\documentclass[aps,prd,reprint,superscriptaddress,preprintnumbers,nofootinbib,floatfix,longbibliography]{revtex4-2}

\usepackage[T1]{fontenc}
\usepackage[utf8]{inputenc}
\usepackage[english]{babel}
\usepackage{amsmath,amsthm,amssymb,amsfonts,mathrsfs,amsbsy,bm}
\usepackage{tensor}
\usepackage{slashed}
\usepackage{bbm}
\usepackage{esint}
\usepackage[a4paper, margin=1.2cm]{geometry}
\usepackage{cancel}
\usepackage{graphicx}
\usepackage{multirow}
\usepackage{array}
\usepackage{booktabs}
\usepackage{makecell}
\usepackage{xcolor}
\colorlet{BLUE}{blue}
\usepackage{tikz}
\usetikzlibrary{quotes,angles,arrows,decorations.markings,decorations.pathmorphing}
\usepackage{hyperref}
\hypersetup{colorlinks=true,breaklinks=true,citecolor=blue,linkcolor=[rgb]{0,0.5,0.9},urlcolor=blue}
\graphicspath{{figures/}}

\newcommand{\be}{\begin{equation}}
\newcommand{\ee}{\end{equation}}
\newcommand{\Be}{\begin{eqnarray}}
\newcommand{\Ee}{\end{eqnarray}}

\newcommand{\mincir}{\raise-3.truept\hbox{\rlap{\hbox{$\sim$}}\raise4.truept\hbox{$<$}\ }}
\newcommand{\magcir}{\raise-3.truept\hbox{\rlap{\hbox{$\sim$}}\raise4.truept\hbox{$>$}\ }}

\providecommand{\U}[1]{}
\newcommand{\ie}{\begin{equation}}
\newcommand{\fe}{\end{equation}}
\newcommand{\se}{\begin{eqnarray}}
\newcommand{\ff}{\end{eqnarray}}

\begin{document}
\emergencystretch=4em


\title{Higher derivative Hořava-Lifshitz fermions: from superconductivity to atomic spectroscopy}

\author{A. A. Ara\'{u}jo Filho}
\email{dilto@fisica.ufc.br}
\affiliation{Departamento de F\'isica, Universidade Federal da Para\'iba, Caixa Postal 5008, 58051--970, Jo\~ao Pessoa, Para\'iba, Brazil.}
\affiliation{Departamento de F\'isica, Universidade Federal de Campina Grande, Caixa Postal 10071, 58429--900 Campina Grande, Para\'iba, Brazil.}
\affiliation{Center for Theoretical Physics, Khazar University, 41 Mehseti Street, Baku, AZ-1096, Azerbaijan.}


\author{K. E. L. de Farias}
\email{klecio.lima@uaf.ufcg.edu.br}
\affiliation{Departamento de F\'isica, Universidade Federal de Campina Grande, Caixa Postal 10071, 58429--900 Campina Grande, Para\'iba, Brazil.}
\affiliation{Centre of Excellence ENSEMBLE3 Sp. z o. o., Wolczynska Str. 133, 01-919, Warsaw, Poland}
\author{N. Heidari}
\email{heidari.n@gmail.com}

\affiliation{Departamento de F\'isica, Universidade Federal de Campina Grande, Caixa Postal 10071, 58429--900 Campina Grande, Para\'iba, Brazil.}
\affiliation{Center for Theoretical Physics, Khazar University, 41 Mehseti Street, Baku, AZ-1096, Azerbaijan.}
\affiliation{School of Physics, Damghan University, Damghan, 3671641167, Iran.}

\author{M. Paganelly}
\email{paganelly.matheus@ifrn.edu.br}
\affiliation{Departamento de F\'isica, Universidade Federal da Para\'iba, Caixa Postal 5008, 58051--970, Jo\~ao Pessoa, Para\'iba, Brazil.}
\affiliation{Instituto Federal de Educação, Ciência e Tecnologia do Estado do Rio Grande do Norte, 59500-000, Macau, Rio Grande do Norte, Brazil.}


\author{A. F. Santos}
\email{alesandroferreira@fisica.ufmt.br}
\affiliation{Programa de Pós-graduação em Física, Instituto de Física, Universidade Federal de Mato Grosso, Cuiabá, Brasil}




\begin{abstract}

We investigate an isotropic Hořava--Lifshitz fermion theory supplemented by a dimension--six spatial derivative operator, preserving first order time evolution while modifying the ultraviolet dispersion relation. Within the regime of validity of the derivative expansion, we develop three complementary consequences of the $z=2$ sector. First, a Bardeen--Cooper--Schrieffer/Nambu--Jona--Lasinio (BCS/NJL) mean field treatment of fermion pairing yields the particle and antiparticle quasiparticle spectra, the finite temperature gap equation, and the gauge invariant criterion for a stable superconducting phase. The modified dispersion alters both the density of states governing the pairing scale and the electromagnetic vertices entering the Meissner response. Second, using the scalar pole denominator as a minimal spin independent exchange kernel, we derive an exponentially damped and oscillatory massive potential and obtain the corresponding Born scattering amplitudes. The quartic momentum dependence renders both the forward and total cross sections finite and produces a large transfer behavior distinct from Rutherford scattering. Finally, we evaluate the induced shifts of the hydrogenic $1S$, $2S$, and $2P$ levels in closed form, retaining the full dependence on the interaction range. In other words, these latter results gives direct spectroscopic constraints on the product of nuclear and leptonic couplings, with muonic atoms, showing the respective sensitivity to the short distance Hořava--Lifshitz sector.

\end{abstract}
 \maketitle


\section{Introduction}
\label{sec:introduction}

Ho\v{r}ava--Lifshitz scaling allows the ultraviolet behavior of a field theory to be modified through higher spatial derivatives while retaining the original order of time evolution. This separation underlies the proposal of gravity at a Lifshitz point and the weighted power-counting formulation of Lorentz-violating matter theories \cite{Horava:2009uw,Anselmi:2008bq,AnselmiTaiuti:2010,Anselmi:2008ry}. Beyond power counting, perturbative renormalizability has been established for projectable Ho\v{r}ava gravity \cite{Barvinsky2016}. More recent analyses of its marginal couplings in $3+1$ dimensions have identified asymptotically free fixed points and trajectories approaching a region with the general-relativistic kinetic structure \cite{BarvinskyKurovSibiryakov2025}.

For fermions, the dynamical exponent does not uniquely determine the spatial kinetic operator. Distinct $z=2$ constructions differ in their Dirac structure, locality, and symmetry properties \cite{Montani:2012cu}, while quantum corrections can modify the relation between fermionic and electromagnetic propagation \cite{AlexandreBrister:2013}. Higher order calculations in four fermion Lifshitz models also show that improved ultraviolet power counting does not by itself suppress Lorentz violation in the infrared \cite{AlexandreBristerHouston2012}. A phenomenological treatment must therefore distinguish the anisotropic kinetic terms from the relevant operators needed to recover ordinary low energy propagation \cite{Iengo:2009ix}. The experimental sensitivities compiled in the 2026 edition of the Lorentz-- and CPT--violation data tables make this matching requirement particularly restrictive \cite{KosteleckyRussell2011}.

Higher dimensional fermionic operators provide a systematic description of departures from relativistic dispersion relations \cite{MyersPospelov:2003,KosteleckyMewes:2013,Maccione:2009ju}. Their observable effects depend on the symmetry of the chosen coefficients and on the interactions used to probe them; a preferred spatial direction and a rotationally invariant deformation are not interchangeable. Gauge couplings introduce a further constraint, since a modified kinetic operator must be accompanied by compatible interaction vertices \cite{KosteleckyLi:2019}. We adopt a rotationally invariant deformation with two additional spatial derivatives and retain first-order time evolution. For $z=2$, the chosen Dirac-type realization involves a fractional Laplacian, so it is understood as a momentum-space effective theory rather than a local polynomial fermion theory.

Pairing provides a direct probe of the resulting density of states. The Bardeen--Cooper--Schrieffer (BCS) mechanism relates an attractive interaction near a Fermi surface to an exponentially generated gap \cite{Bardeen:1957mv}, whereas the Nambu--Jona-Lasinio (NJL) construction formulates dynamical symmetry breaking through a four-fermion interaction \cite{NambuJonaLasinio1961,Tong:2023krn}. In the Lifshitz setting, a $z=3$ four-fermion theory can exhibit asymptotic freedom and dynamical mass generation in a large-species expansion \cite{DharMandalWadia2009}. Those results concern a different scaling regime and condensate from the finite-density pairing considered here. We use an NJL-type contact interaction in a BCS pairing channel and retain both particle and antiparticle excitations. Magnetic screening requires a separate response calculation: gauge invariance ties the fermionic vertices to fluctuations of the order parameter, and a nonzero gap alone does not establish a stable superconducting state \cite{Nambu:1960tm,GuoChienHe:2012}.

Static interactions and elastic scattering test additional information that is not contained in the dispersion relation alone. Recent studies have examined modifications of electron scattering in noncommutative geometry and bumblebee gravity, including changes in screening, angular dependence, and effective coupling strengths \cite{Touati:2025,AraujoFilho:2026yaj,AraujoFilho:2026oqc}. In each case, an interaction channel is needed to connect propagation to a source-dependent potential. The same distinction is essential for a Lifshitz fermion: its matrix-valued propagator cannot be replaced by a scalar exchange amplitude without an additional assumption. Once that assumption is fixed, the Born construction must use the appropriate incident flux and external spinors \cite{Born:1926,LippmannSchwinger:1950,Mott:1929}.

Precision spectroscopy constrains weak interactions through differences of bound-state matrix elements. Measurements in muonic hydrogen, deuterium, and helium have extended this sensitivity to orbital scales much smaller than those of electronic atoms \cite{Pohl2010,Antognini2013,Pohl2016,Krauth2021}, with nuclear structure setting an important part of the theoretical uncertainty \cite{Pachucki2024}. Molecular ions and antiprotonic helium probe additional pair separations and coupling combinations \cite{Salumbides2014,Germann2021}, while muonium avoids a composite nuclear source \cite{MuMASS2022}. Recent isotope-shift measurements in ytterbium and calcium further constrain electron--neutron interactions while exposing the role of higher-order nuclear contributions \cite{Door2025,Wilzewski2025}. These advances also sharpen the interpretation of spectroscopic limits: nuclear radii, fundamental constants, and possible new interactions must be constrained with independent inputs or included in a common fit \cite{Delaunay2023}. Bounds derived for a Yukawa potential cannot be transferred unchanged to a kernel with a different radial profile.

In this work, we study the isotropic spatial factor 
\begin{equation} 
F_z(p) = \frac{p^{z-1}}{\Lambda_{\mathrm{HL}}^{z-1}}\left(1-\eta p^2\right), \qquad \eta = \frac{\bar{\alpha}}{M_P^2}, 
\nonumber
\end{equation} 
within a momentum domain below the effective theory cutoff and satisfying $|\eta|p^2\ll1$. We derive the quasiparticle spectrum and finite temperature gap equation, then construct the gauge-covariant response needed to formulate the Meissner criterion. For $z=2$, we separately introduce a spin independent exchange kernel with the same spectral denominator, derive its massive and massless potentials, and calculate density coupled electron scattering with the group velocity and spinor overlaps retained. The pairing interaction and exchange strengths remain independent inputs; neither is determined by the free dispersion relation. For the atomic application, we evaluate the hydrogenic $1S$, $2S$, and $2P$ shifts without expanding in the ratio of interaction range to Bohr radius. This comparison assumes that conventional atomic kinematics has been recovered by infrared matching and that the static kernel enters as an additional weak interaction. The resulting expressions allow spectroscopic constraints to be imposed with explicit nuclear radius inputs, source distributions, and effective theory uncertainties.

Sections~\ref{sec2} and \ref{sec3} define the fermionic scaling and its higher-spatial-derivative extension. Section~\ref{sec:HL_BCS_superconductivity} develops the pairing and electromagnetic response. Sections~\ref{sec:HL_effective_potential} and \ref{sec:HL_electron_scattering} derive the static potential and scattering cross sections, and Sec.~\ref{sec:atomic-bounds} examines the spectroscopic constraints.


\section{Ho\v{r}ava--Lifshitz scaling for fermions }
\label{sec2}

Ho\v{r}ava--Lifshitz theories distinguish the scaling of space and time according to
\begin{equation} \boldsymbol{x}\rightarrow b\,\boldsymbol{x}, \qquad t_{\mathrm{L}}\rightarrow b^z t_{\mathrm{L}}, \label{eq:HL_anisotropic_scaling} \end{equation}
where $b>0$, $z$ denotes the dynamical critical exponent, and $t_{\mathrm{L}}$ represents time in weighted units. The relativistic scaling is recovered for $z=1$, while $z>1$ permits higher spatial derivatives without introducing higher time derivatives. This anisotropy underlies the improved ultraviolet power counting of Lifshitz-type theories \cite{Horava:2009uw,Anselmi:2008bq,AnselmiTaiuti:2010}. Its implementation in the fermionic sector requires particular care because the conventional Dirac operator is first order in both space and time \cite{Montani:2012cu,AlexandreBrister:2013}.

The ordinary Dirac Lagrangian is
\begin{equation} 
\mathcal{L}_{\mathrm{D}} = \bar{\psi}\left(i\gamma^0\partial_0+i\gamma^i\partial_i-m\right)\psi. 
\label{eq:HL_standard_Dirac_Lagrangian} 
\end{equation}
Under the transformation in Eq.~\eqref{eq:HL_anisotropic_scaling}, the weighted momentum dimensions are
\begin{equation} 
[x^i]_{\mathrm{w}}=-1, \quad [t_{\mathrm{L}}]_{\mathrm{w}}=-z, \quad [\partial_i]_{\mathrm{w}}=1, \quad [\partial_{t_{\mathrm{L}}}]_{\mathrm{w}}=z. 
\label{eq:HL_weighted_dimensions} 
\end{equation}
In $d$ spatial dimensions, invariance of the temporal kinetic term,
\begin{equation} S_{\mathrm{time}} = \int\mathrm{d}t_{\mathrm{L}}\,\mathrm{d}^dx\, \bar{\psi}i\gamma^0\partial_{t_{\mathrm{L}}}\psi, 
\label{eq:HL_temporal_action} 
\end{equation}
requires
\begin{equation} 
2[\psi]_{\mathrm{w}} + z - (d+z) = 0, \quad [\psi]_{\mathrm{w}} =[\bar{\psi}]_{\mathrm{w}} = \frac{d}{2}. 
\label{eq:HL_fermion_weight} 
\end{equation}
Thus, in three spatial dimensions, $[\psi]_{\mathrm{w}}=3/2$, independently of $z$. The coefficient of the bilinear $\bar{\psi}\psi$ has weighted dimension $[m_{\mathrm{L}}]_{\mathrm{w}} = z$ and constitutes a relevant deformation of the scale-invariant theory.

A rotationally invariant spatial operator with weighted dimension $z$ may be defined as
\begin{equation} 
\mathcal{D}_z\equiv i\gamma^i\partial_i(-\nabla^2)^{(z-1)/2}, 
\label{eq:HL_Dz_definition} 
\end{equation}
where $\nabla^2 = \delta^{ij} \partial_i \partial_j$. In momentum space,
\begin{equation} 
\mathcal{D}_z\longrightarrow\gamma^ip_i\,p^{z-1}, \quad p = |\boldsymbol{p}|. 
\label{eq:HL_Dz_momentum} 
\end{equation}
This form avoids ambiguous expressions such as $(i \partial_i)^z$, in which the same spatial index would appear more than twice.

The corresponding ultraviolet action in three spatial dimensions is
\begin{equation} 
S_{\mathrm{L}} = \int\mathrm{d}t_{\mathrm{L}}\,\mathrm{d}^3x\,\bar{\psi}\left(i\gamma^0\partial_{t_{\mathrm{L}}}+\mathcal{D}_z -m_{\mathrm{L}}\right)\psi. 
\label{eq:HL_unrescaled_fermion_action} 
\end{equation}
The temporal and spatial kinetic operators both have weighted dimension $z$. Under Eq.~\eqref{eq:HL_anisotropic_scaling}, the massless action is invariant when the fermion transforms as
\begin{equation} 
\psi\rightarrow b^{-3/2}\psi, \quad \bar{\psi}\rightarrow b^{-3/2}\bar{\psi}. 
\label{eq:HL_fermion_scaling} 
\end{equation}

To express the theory in conventional energy units, we introduce the Ho\v{r}ava--Lifshitz scale $\Lambda_{\mathrm{HL}}$ and define
\begin{equation} 
t_{\mathrm{L}} = \frac{t}{\Lambda_{\mathrm{HL}}^{z-1}}, \quad \partial_{t_{\mathrm{L}}} = \Lambda_{\mathrm{HL}}^{z-1}\partial_t, \qquad m = \frac{m_{\mathrm{L}}}{\Lambda_{\mathrm{HL}}^{z-1}}. 
\label{eq:HL_time_mass_rescaling} 
\end{equation}
The integration measure transforms as
\begin{equation} 
\mathrm{d}t_{\mathrm{L}}\,\mathrm{d}^3x = \frac{\mathrm{d}t\,\mathrm{d}^3x}{\Lambda_{\mathrm{HL}}^{z-1}}. 
\label{eq:HL_measure_rescaling} 
\end{equation}
The Jacobian is canceled by the rescaling of the temporal derivative, so no additional redefinition of the fermion field is required. The resulting action is
\begin{equation} 
S_0 = \int\mathrm{d}t\, \mathrm{d}^3x\, \bar{\psi}\left[i\gamma^0\partial_0+\frac{i\gamma^i\partial_i(-\nabla^2)^{(z-1)/2}}{\Lambda_{\mathrm{HL}}^{z-1}}-m\right]\psi. 
\label{eq:HL_rescaled_fermion_action} 
\end{equation}
In conventional units, $\psi$ has mass dimension $3/2$, whereas $m$ and $\Lambda_{\mathrm{HL}}$ have mass dimension one.

Adopting the momentum-space convention $i\partial_0\rightarrow p_0$ and $i\partial_i\rightarrow p_i$, the inverse propagator becomes
\begin{equation} 
S_0^{-1}(p_0,\boldsymbol{p}) = \gamma^0 p_0 + \gamma^ip_i\frac{p^{z-1}}{\Lambda_{\mathrm{HL}}^{z-1}} - m. 
\label{eq:HL_free_inverse_propagator} 
\end{equation}
Multiplication by the conjugate Dirac operator gives
\begin{equation} 
\begin{split}
&\left(\gamma^0 p_0 + \gamma^i p_i\frac{p^{z-1}}{\Lambda_{\mathrm{HL}}^{z-1}} -m\right) \\
& \times \left(\gamma^0 p_0+\gamma^ip_i\frac{p^{z-1}}{\Lambda_{\mathrm{HL}}^{z-1}} + m\right) = p_0^2 - m^2 - \frac{p^{2z}}{\Lambda_{\mathrm{HL}}^{2z-2}}. 
\label{eq:HL_free_kernel_product} 
\end{split}
\end{equation}
The poles are therefore determined by
\begin{equation} 
p_0^2 = m^2 + \frac{p^{2z}}{\Lambda_{\mathrm{HL}}^{2z-2}}, 
\label{eq:HL_free_dispersion_relation} 
\end{equation}
and the positive energy scenario is
\begin{equation} 
E(p) = \sqrt{m^2 + \frac{p^{2z}}{\Lambda_{\mathrm{HL}}^{2z-2}}}. 
\label{eq:HL_free_positive_energy} 
\end{equation}
In the regime $p^z/\Lambda_{\mathrm{HL}}^{z - 1}\gg m$, the energy behaves as $E(p)\sim p^z/\Lambda_{\mathrm{HL}}^{z-1}$. Setting $z=1$ reproduces the relativistic Dirac dispersion, as we should expect.

For positive odd integer $z$, the exponent $(z-1)/2$ is an integer and $\mathcal{D}_z$ is a local differential operator. For even integer $z$, the same Dirac type realization contains a fractional power of the Laplacian, defined spectrally by
\begin{equation} 
(-\nabla^2)^\nu e^{i\boldsymbol{p}\cdot\boldsymbol{x}} = p^{2\nu}e^{i\boldsymbol{p}\cdot\boldsymbol{x}}. 
\label{eq:HL_fractional_Laplacian_definition} 
\end{equation}
In particular, the $z=2$ model employed below contains $(-\nabla^2)^{1/2}$ and is therefore pseudodifferential in position space, although its momentum space formulation is well defined \cite{Montani:2012cu}. Local $z=2$ fermionic theories may instead be constructed with a Laplacian multiplying the identity in spinor space, but they possess a different Dirac structure and do not yield the numerator adopted here \cite{AlexandreBrister:2013}.

At fixed $z>1$, Eq.~\eqref{eq:HL_rescaled_fermion_action} defines a pure Lifshitz model rather than a complete interpolation between relativistic and anisotropic regimes. Such an interpolation requires the inclusion of a relevant spatial Dirac operator linear in momentum, whose coefficient is fixed by infrared matching \cite{AnselmiTaiuti:2010,AlexandreBrister:2013}. Indeed, for $z=2$ and $m>0$, the low-momentum expansion of Eq.~\eqref{eq:HL_free_positive_energy} is
\begin{equation} 
E(p) = m + \frac{p^4}{2m\Lambda_{\mathrm{HL}}^2} + \mathcal{O}\left(\frac{p^8}{m^3\Lambda_{\mathrm{HL}}^4}\right), 
\label{eq:HL_z2_low_momentum} 
\end{equation}
which contains a quartic kinetic term rather than the quadratic kinetic energy of an ordinary nonrelativistic fermion. In the present analysis, Eq.~\eqref{eq:HL_rescaled_fermion_action} is taken as the defining isotropic momentum space model.


\section{Higher spatial derivative extension of the fermionic sector }
\label{sec3}

Higher dimensional Lorentz--violating operators provide a systematic description of suppressed corrections to fermion propagation \cite{MyersPospelov:2003,KosteleckyMewes:2013,KosteleckyLi:2019}. A dimension--six fermionic operator constructed from a constant, dimensionless background vector $n^\mu$ is
\begin{equation} 
\delta\mathcal{L}_n = \frac{i}{M_P^2}\bar{\psi}(n\cdot\partial)^3\not{n}\left(\bar{\alpha}_LP_L +\bar{\alpha}_RP_R\right)\psi, 
\label{eq:HL_fixed_vector_operator} 
\end{equation}
where $M_P$ is the suppression scale and $\bar{\alpha}_L$ and $\bar{\alpha}_R$ are real dimensionless coefficients. We use $n\cdot\partial = n^\mu\partial_\mu$, $\not n = \gamma^\mu n_\mu$, and the metric $g_{\mu\nu} = \operatorname{diag}(1,-1,-1,-1)$. The chiral projectors are
\begin{equation} 
P_L = \frac{1-\gamma^5}{2}, \quad P_R = \frac{1 + \gamma^5}{2}. 
\label{eq:HL_chiral_projectors} 
\end{equation}
Defining $\bar{\alpha}_V = (\bar{\alpha}_R + \bar{\alpha}_L)/2$ and $\bar{\alpha}_A =(\bar{\alpha}_R - \bar{\alpha}_L)/2$, we obtain
\begin{equation} 
\bar{\alpha}_LP_L + \bar{\alpha}_RP_R = \bar{\alpha}_V + \bar{\alpha}_A\gamma^5. 
\label{eq:HL_vector_axial_decomposition} 
\end{equation}
Equal chiral coefficients, $\bar{\alpha}_L = \bar{\alpha}_R = \bar{\alpha}_n$, eliminate the axial contribution and give
\begin{equation} 
\delta\mathcal{L}_n = \frac{i\bar{\alpha}_n}{M_P^2}\bar{\psi}(n\cdot\partial)^3\not{n}\psi. 
\label{eq:HL_vectorlike_fixed_vector_operator}
\end{equation}
For unequal coefficients, the axial operator must be retained and the vector like dispersion derived below does not apply.

A purely spacelike background $n^\mu = (0,\boldsymbol{n})$ satisfies $n\cdot\partial=\boldsymbol{n}\cdot\boldsymbol{\nabla}$ and $\not{n} = -\boldsymbol{\gamma}\cdot\boldsymbol{n}$. In this manner,
\begin{equation} 
\delta\mathcal{L}_n = - \frac{i\bar{\alpha}_n}{M_P^2}\bar{\psi}(\boldsymbol{n}\cdot\boldsymbol{\nabla})^3(\boldsymbol{\gamma}\cdot\boldsymbol{n})\psi. 
\label{eq:HL_spacelike_fixed_vector_operator} 
\end{equation}
With the Fourier convention $i\partial_\mu\rightarrow p_\mu$, the correction to the quadratic Dirac kernel is
\begin{equation} 
\delta\mathcal{K}_n(p) = - \frac{\bar{\alpha}_n}{M_P^2}(n\cdot p)^3\not{n}. 
\label{eq:HL_fixed_vector_momentum_kernel} 
\end{equation}
Its dependence on $\boldsymbol{n}\cdot\boldsymbol{p}$ selects a spatial direction. No choice of a single nonzero $\boldsymbol{n}$ converts this tensor structure into the rotationally invariant contraction $\gamma^ip_i p^2$.

In the preferred Lifshitz frame, we instead choose the isotropic dimension--six operator
\begin{equation} 
\delta\mathcal{L}_{\mathrm{iso}} = -\frac{\bar{\alpha}}{M_P^2}\bar{\psi}\,i\gamma^i\partial_i(-\nabla^2)\psi, 
\label{eq:HL_isotropic_dimension_six_operator}
\end{equation}
whose momentum space form is
\begin{equation} 
\delta\mathcal{L}_{\mathrm{iso}}\longrightarrow-\frac{\bar{\alpha}}{M_P^2}\bar{\psi}\,\gamma^ip_i p^2\psi. 
\label{eq:HL_isotropic_dimension_six_momentum} 
\end{equation}
The real coefficient $\bar{\alpha}$ is independent of $\bar{\alpha}_n$, since the isotropic and fixed vector operators describe different deformations. The overall minus sign fixes the convention for $\bar{\alpha}$.

We extend Eq.~\eqref{eq:HL_isotropic_dimension_six_operator} to the Lifshitz model by inserting two additional spatial derivatives into the operator $\mathcal{D}_z$ defined in Sec.~\ref{sec2}. Thereby,
\begin{equation} 
\begin{split}
& \mathcal{D}_{z + 2} \equiv \, \mathcal{D}_z(-\nabla^2) = i\gamma^i\partial_i(-\nabla^2)^{(z+1)/2}, \\
& \mathcal{D}_{z+2}\longrightarrow\gamma^ip_i p^{z+1}. 
\label{eq:HL_Dzplus2_definition} 
\end{split}
\end{equation}
Using the same time rescaling as in the free action gives
\begin{equation} 
\delta S_{\mathrm{HD}} = -\int\mathrm{d}t\,\mathrm{d}^3x\,\bar{\psi}\left[\frac{\bar{\alpha}}{M_P^2}\frac{\mathcal{D}_{z + 2}}{\Lambda_{\mathrm{HL}}^{z-1}}\right]\psi.
\label{eq:HL_high_derivative_action} 
\end{equation}
This is a chosen extension compatible with spatial rotations; anisotropic scaling alone does not determine it. In ordinary mass dimensions, the bilinear $\bar{\psi}\mathcal{D}_{z+2}\psi$ has dimension $z + 5$, while its coefficient has dimension $-(z+1)$, leaving the Lagrangian density with dimension four. The undivided bilinear is  dimension--six for $z=1$ and dimension seven for $z = 2$.

Introducing $\eta\equiv\bar{\alpha}/M_P^2$, with ordinary mass dimension $-2$, the quadratic action becomes
\begin{equation} 
S = \int\mathrm{d}t\,\mathrm{d}^3x\,\bar{\psi}\left[i\gamma^0\partial_0+\frac{\mathcal{D}_z}{\Lambda_{\mathrm{HL}}^{z-1}}\left(1-\eta(-\nabla^2)\right)-m\right]\psi. 
\label{eq:HL_complete_fermion_action} 
\end{equation}
Its momentum space Dirac kernel is
\begin{equation} 
\mathcal{K}_z(p_0,\boldsymbol{p}) = \gamma^0p_0+\gamma^ip_iF_z(p) - m, 
\label{eq:HL_complete_inverse_propagator} 
\end{equation}
where
\begin{equation} 
F_z(p) = \frac{p^{z-1}}{\Lambda_{\mathrm{HL}}^{z-1}}\left(1-\eta p^2\right). 
\label{eq:HL_Fz_definition} 
\end{equation}
The Clifford algebra yields
\begin{widetext}
\begin{equation} 
\left[\gamma^0 p_0 + \gamma^i p_i F_z(p) -m \right]\left[\gamma^0 p_0 + \gamma^i p_i F_z(p) + m\right] = \left[p_0^2 - m^2 -p^2 F_z(p)^2\right]\mathbbm{1}_4. 
\label{eq:HL_propagator_denominator_derivation} 
\end{equation}
\end{widetext}
The Feynman propagator, obtained as $i\mathcal{K}_z^{-1}$ with the causal prescription, is 
\begin{equation} 
S_F(p_0,\boldsymbol{p}) = i\frac{\gamma^0 p_0 + \gamma^i p_i F_z(p) + m}{p_0^2 - m^2-p^2 F_z(p)^2 + i0^+}. 
\label{eq:HL_complete_fermion_propagator} 
\end{equation}
Its poles occur at $p_0=\pm E(p)$, with
\begin{equation} 
E(p) = \sqrt{m^2 + \frac{p^{2z}}{\Lambda_{\mathrm{HL}}^{2z - 2}}\left(1 - \eta p^2\right)^2}. 
\label{eq:HL_complete_dispersion_relation} 
\end{equation}

Eq.~\eqref{eq:HL_complete_dispersion_relation} is exact for the displayed quadratic truncation. Its terms quadratic in $\eta$ do not constitute a complete prediction at the next order in the effective expansion, where omitted operators may also contribute. To first order,
\begin{equation} 
E(p)^2 = m^2 + \frac{p^{2z}}{\Lambda_{\mathrm{HL}}^{2z-2}}\left(1-2\eta p^2\right) + \mathcal{O}(\eta^2). 
\label{eq:HL_dispersion_relation_expanded} 
\end{equation}
Writing $E_0(p) = \sqrt{m^2 + p^{2z}/\Lambda_{\mathrm{HL}}^{2z - 2}}$, the corresponding energy shift is
\begin{equation} 
E(p) = E_0(p) - \eta\frac{p^{2z+2}}{\Lambda_{\mathrm{HL}}^{2z - 2}E_0(p)}+\mathcal{O}(\eta^2). 
\label{eq:HL_energy_shift_expanded} 
\end{equation}
Notice that at fixed nonzero momentum, positive $\eta$ lowers the energy, whereas negative $\eta$ raises it within the perturbative regime.

The relative deformation of the spatial kinetic operator is controlled by $|\eta|p^2\ll1$. All momenta must also remain below the cutoff of the underlying effective description; the smallness of $|\eta|p^2$ alone does not determine that cutoff. For $\eta>0$, the factor $1-\eta p^2$ vanishes at $p^2=\eta^{-1}$, where the retained correction is as large as the leading term. The unexpanded expression then gives $E=|m|$, but this finite-momentum feature lies outside the regime pointed out above and cannot be regarded as a controlled prediction.

The spatial deformation leaves the action first order in time. For real parameters, the free Hamiltonian $H(\boldsymbol{p}) = \boldsymbol{\alpha}\cdot\boldsymbol{p}\,F_z(p) + \beta m$, with $\alpha^i=\gamma^0\gamma^i$ and $\beta = \gamma^0$, is Hermitian. The particle and antiparticle case each retain their twofold spin degeneracy, and the higher spatial derivatives introduce no additional frequency poles.

For the $z=2$ sector used below,
\begin{equation} 
F_2(p) = \frac{p}{\Lambda_{\mathrm{HL}}}\left(1-\eta p^2\right), 
\label{eq:HL_F2_final} 
\end{equation}
and
\begin{equation} 
E(p) = \sqrt{m^2+\frac{p^4}{\Lambda_{\mathrm{HL}}^2}\left(1 - \eta p^2\right)^2}. 
\label{eq:HL_z2_dispersion_relation} 
\end{equation}
To the retained order, the correction to the quartic dispersion is $-2\eta p^6/\Lambda_{\mathrm{HL}}^2$ in $E(p)^2$.

\begin{figure}[!htbp]
\centering
\makebox[\columnwidth][l]{
   \hspace*{-0.7cm}
   \includegraphics[width=1.1\columnwidth]{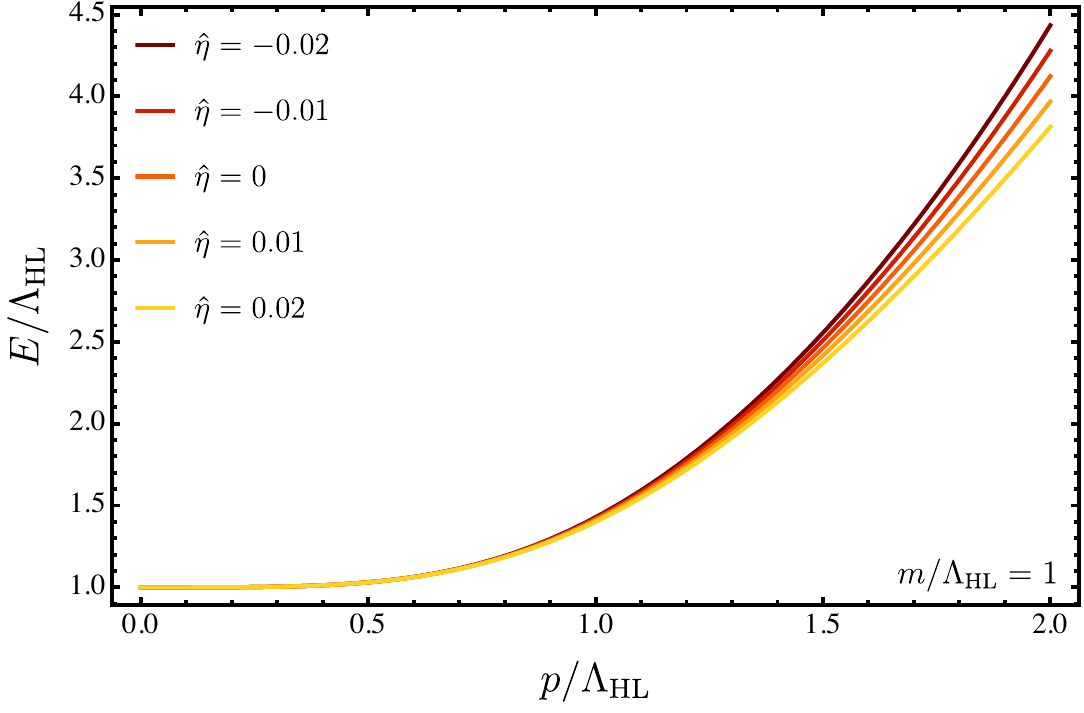}}
\caption{Dimensionless positive energy dispersion, $E/\Lambda_{\mathrm{HL}}$, as a function of $p/\Lambda_{\mathrm{HL}}$ for $m/\Lambda_{\mathrm{HL}} = 1$ and $\widehat{\eta} = \{-0.02,-0.01,0,0.01,0.02\}$.}
\label{fig:hl-dispersion}
\end{figure}

Fig.~\ref{fig:hl-dispersion} illustrates the deformation of the positive energy segment in the $z=2$ sector. The curves remain nearly indistinguishable at low momentum, where the mass term dominates, and separate progressively once the quartic contribution becomes relevant.  In agreement with Eq.~(\ref{eq:HL_z2_dispersion_relation}), positive $\widehat{\eta}$ lowers the energy at fixed momentum, whereas negative $\widehat{\eta}$ raises it.  The displayed interval also satisfies $|\widehat{\eta}|(p/\Lambda_{\mathrm{HL}})^2\leq 0.08$, so that the comparison does not approach the breakdown of the derivative expansion.

\section{Mean field pairing and electromagnetic response}
\label{sec:HL_BCS_superconductivity}

We examine homogeneous pairing of a single Dirac species at chemical potential $\mu\geq0$. The normal-state spectrum is
\begin{equation} 
E(p) = \sqrt{m^2+p^2F(p)^2}, \quad F(p) = \frac{p^{z-1}}{\Lambda_{\mathrm{HL}}^{z-1}}\left(1-\eta p^2\right), 
\label{eq:HL_BCS_single_particle_energy} 
\end{equation}
where $\eta=\bar{\alpha}/M_P^2$ and $p=|\boldsymbol{p}|$. We define
\begin{equation} 
\xi_-(p) = E(p) - \mu, \quad \xi_+(p) = E(p) + \mu, 
\label{eq:HL_BCS_particle_antiparticle_energies} 
\end{equation}
which describe the particle and antiparticle sectors, respectively. Both are retained in the mean field calculation.

An attractive interaction in the rotationally scalar, even parity pairing channel is
\begin{equation} 
\mathcal{L}_{\mathrm{int}} = G\left(\bar{\psi}i\gamma_5C\bar{\psi}^{\,T}\right)\left(\psi^TCi\gamma_5\psi\right), \quad G>0,
\label{eq:HL_BCS_four_fermion} 
\end{equation}
where $G>0$ is the coupling constant. In conventional units, $G$ has mass dimension $-2$ and is an independent interaction parameter. We use $C = i\gamma^2\gamma^0$, with $C^T=-C$ and $C^{-1}\gamma^\mu C = -(\gamma^\mu)^T$. The matrix $Ci\gamma_5$ is antisymmetric, so the bilinear is compatible with Fermi statistics. This contact interaction provides a mean-field realization of Bardeen--Cooper--Schrieffer pairing \cite{Bardeen:1957mv,Nambu:1960tm}.

The order parameter is defined by
\begin{equation} 
\Delta = -2G\left\langle\psi^TCi\gamma_5\psi\right\rangle, \qquad \Delta^* =  - 2G\left\langle\bar{\psi}i\gamma_5C\bar{\psi}^{\,T}\right\rangle. 
\label{eq:HL_BCS_gap_definition} 
\end{equation}
Neglecting terms quadratic in fluctuations about these expectation values gives
\begin{equation} 
\mathcal{L}_{\mathrm{int}}^{\mathrm{MF}} = -\frac{|\Delta|^2}{4G}-\frac{1}{2}\left[\Delta^*\psi^TCi\gamma_5\psi+\Delta\bar{\psi}i\gamma_5C\bar{\psi}^{\,T}\right]. 
\label{eq:HL_BCS_mean_field_interaction} 
\end{equation}
Under $\psi\rightarrow e^{i\chi}\psi$, the condensate transforms as $\Delta\rightarrow e^{2i\chi}\Delta$. A nonzero expectation value therefore breaks the global particle number symmetry; for fermion charge $e$, the pair carries charge $2e$.

Introducing $\psi_C = C\bar{\psi}^{\,T}$ and $\bar{\psi}_C = \psi^TC$, we write
\begin{equation} 
\Psi = \begin{pmatrix}\psi\\ \psi_C\end{pmatrix}, \qquad \overline{\Psi}=\left(\bar{\psi},\bar{\psi}_C\right). 
\label{eq:HL_BCS_Nambu_spinor} 
\end{equation}
For $P=(i\omega_n,\boldsymbol{p})$, with $\omega_n=(2n+1)\pi T$, define the normal Dirac kernels
\begin{equation} 
\mathcal{K}_{\pm}(P) = \gamma^0(i\omega_n\pm\mu)+\gamma^ip_iF(p) - m. 
\label{eq:HL_BCS_normal_kernels} 
\end{equation}
The inverse Nambu--Gor'kov propagator consistent with Eq.~\eqref{eq:HL_BCS_gap_definition} is
\begin{equation} 
\mathcal{G}^{-1}(P) = \begin{pmatrix}\mathcal{K}_+(P)& - \Delta i\gamma_5\\ -\Delta^*i\gamma_5&\mathcal{K}_-(P)\end{pmatrix}. 
\label{eq:HL_BCS_inverse_Nambu_propagator} 
\end{equation}
A global phase choice makes the homogeneous gap real and non-negative. The determinant then factorizes as
\begin{equation} 
\det\mathcal{G}^{-1}(P) = \left[\omega_n^2+\mathcal{E}_-(p)^2\right]^2\left[\omega_n^2 + \mathcal{E}_+(p)^2\right]^2, 
\label{eq:HL_BCS_Nambu_determinant}
\end{equation}
where
\begin{equation} 
\mathcal{E}_-(p) = \sqrt{\xi_-(p)^2+\Delta^2}, \qquad \mathcal{E}_+(p) = \sqrt{\xi_+(p)^2+\Delta^2}. 
\label{eq:HL_BCS_quasiparticle_branches} 
\end{equation}
The two branches retain the spin degeneracy of the normal spectrum. The antiparticle branch can be integrated out when its excitation scale is well above the pairing energies, but its contribution to the effective interaction must be included in the matching.

We regulate the momentum integrals according to
\begin{equation} 
\int_{\boldsymbol{p}}\equiv\int_{|\boldsymbol{p}|<\Lambda_{\mathrm{UV}}}\frac{\mathrm{d}^3p}{(2\pi)^3}, \quad |\eta|\Lambda_{\mathrm{UV}}^2\ll1, 
\label{eq:HL_BCS_regulated_integral} 
\end{equation}
with $\Lambda_{\mathrm{UV}}$ below the cutoff of the underlying effective description. Integrating out the fermions gives the grand-potential density relative to the normal state,
\begin{widetext}
\begin{equation} 
\Omega(\Delta,T,\mu) -\Omega(0,T,\mu)=\frac{\Delta^2}{4G}-\frac{\nu}{2}\sum_{s=\pm}\int_{\boldsymbol{p}}\left\{\mathcal{E}_s(p) -|\xi_s(p)|+2T\ln\left[\frac{1+e^{-\mathcal{E}_s(p)/T}}{1+e^{-|\xi_s(p)|/T}}\right]\right\}, 
\label{eq:HL_BCS_thermodynamic_potential} 
\end{equation}
\end{widetext}
where $\nu = 2$ counts the spin states of one Dirac species. The prefactor includes the factor $1/2$ required by Nambu doubling. All thermodynamic derivatives below are taken at fixed $\mu$. At fixed density, the gap equation must be supplemented by the number equation $n = -\partial\Omega/\partial\mu$, using the full grand potential.

For a nonvanishing gap, the stationary condition $\partial\Omega/\partial\Delta=0$ yields
\begin{equation} 
\frac{1}{2G} = \frac{\nu}{2}\sum_{s = \pm}\int_{\boldsymbol{p}}\frac{\tanh[\mathcal{E}_s(p)/(2T)]}{\mathcal{E}_s(p)}. 
\label{eq:HL_BCS_gap_equation} 
\end{equation}
At zero temperature,
\begin{equation} 
\frac{1}{2G} = \frac{\nu}{2}\sum_{s=\pm}\int_{\boldsymbol{p}}\frac{1}{\sqrt{[E(p)+s\mu]^2+\Delta_0^2}}, \quad \Delta_0\equiv\Delta(T=0). 
\label{eq:HL_BCS_zero_temperature_gap} 
\end{equation}
For a continuous transition, the mean field critical temperature follows from $\Delta \rightarrow 0$,
\begin{equation} 
\frac{1}{2G} = \frac{\nu}{2}\sum_{s = \pm}\int_{\boldsymbol{p}}\frac{\tanh\left(|E(p)+s\mu|/(2T_c)\right)}{|E(p)+s\mu|}. 
\label{eq:HL_BCS_critical_temperature} 
\end{equation}

The ultraviolet sensitivity follows directly from the dispersion. At $\eta = 0$, in the regime where the mass is negligible,
\begin{equation} 
\int^{\Lambda_{\mathrm{UV}}}\frac{p^2\,\mathrm{d}p}{E(p)}\sim\Lambda_{\mathrm{HL}}^{z-1}\int^{\Lambda_{\mathrm{UV}}}\mathrm{d}p\,p^{2-z}. 
\label{eq:HL_BCS_UV_power_counting}
\end{equation}
The integral is power law divergent for $z<3$, logarithmically divergent for $z=3$, and ultraviolet convergent for $z>3$. In particular, the $z=2$ gap equation retains linear cutoff sensitivity. Expanding the higher derivative correction adds a contribution proportional to $\eta\Lambda_{\mathrm{HL}}\Lambda_{\mathrm{UV}}^3$ in this case. Its relative size remains small only when Eq.~\eqref{eq:HL_BCS_regulated_integral} holds. In this manner, the unexpanded dispersion therefore cannot be used to claim ultraviolet convergence outside the effective regime.

At $\mu = 0$, the zero-temperature onset of pairing is determined by
\begin{equation} 
\frac{1}{2G_c}=\nu\int_{\boldsymbol{p}}\frac{1}{E(p)}, 
\label{eq:HL_BCS_zero_density_critical_coupling}
\end{equation}
provided the integral is infrared finite. This condition holds for $m > 0$ and, in the massless theory, for $z < 3$. For $m = 0$ and $z \geq 3$, the infrared integral diverges and the threshold vanishes within the mean field model, even though no Fermi surface of finite radius is present.

For $\mu > m$, a normal state Fermi momentum satisfies $E(p_F) = \mu$, provided $p_F$ lies inside the regulated momentum domain. Its velocity is
\begin{equation} 
v_F = \left.\frac{\mathrm{d}E(p)}{\mathrm{d}p}\right|_{p = p_F} = \left.\frac{p F(p)[F(p) + p F'(p)]}{E(p)}\right|_{p = p_F}, 
\label{eq:HL_BCS_Fermi_velocity} 
\end{equation}
with
\begin{equation} 
F'(p) = \frac{p^{z-2}}{\Lambda_{\mathrm{HL}}^{z-1}}\left[(z - 1) -( z + 1)\eta p^2\right]. 
\label{eq:HL_BCS_F_derivative} 
\end{equation}
For $z\geq1$, the dispersion is monotonic throughout $|\eta|p^2\ll1$, so the Fermi surface is unique within the controlled regime. Additional surfaces obtained by extending the polynomial factor to large momentum are outside this approximation. The density of states, including both spin states, is
\begin{equation} 
N(0) = \frac{\nu p_F^2}{2\pi^2|v_F|}.
\label{eq:HL_BCS_density_of_states} 
\end{equation}

In the weak coupling BCS regime, pairing is concentrated in a shell $|\xi_-(p)|<\omega_c$ lying well inside the regulated band. Let $G_{\mathrm{BCS}}$ denote the attraction matched to this shell after integrating out the remaining modes, including the antiparticle sector.

\begin{figure}[!htbp]
\centering
\makebox[\columnwidth][l]{
   \hspace*{-1cm}
   \includegraphics[width=1.1\columnwidth]{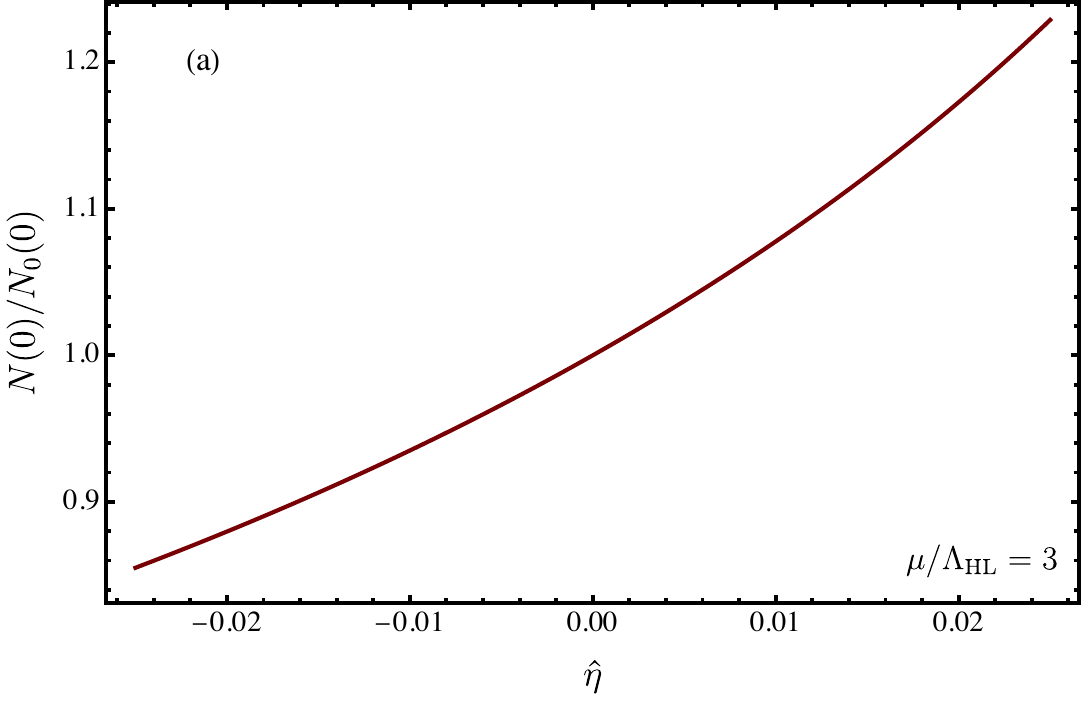}}
\caption{Density of states at the Fermi surface, normalized by its undeformed value, as a function of $\widehat{\eta}$ for $\mu/\Lambda_{\mathrm{HL}}=3$, $m/\Lambda_{\mathrm{HL}}=1$, and spin degeneracy $\nu=2$.}
\label{fig:hl-density-of-states}
\end{figure}

The spectral origin of the pairing correction is shown in Fig.~\ref{fig:hl-density-of-states}. At fixed chemical potential, the normalized density of states grows monotonically with $\widehat{\eta}$. This behavior follows from the simultaneous displacement of the Fermi momentum and reduction of the corresponding group velocity. Although the deformation is perturbatively small, the variation of $N(0)$ is already appreciable over the displayed range and cannot be omitted from the pairing scale.

For an approximately constant density of states,
\begin{equation} 
\Delta_0 \simeq 2 \omega_c \exp \left[-\frac{1}{2G_{\mathrm{BCS}}N(0)}\right]. 
\label{eq:HL_BCS_weak_coupling_gap} 
\end{equation}

\begin{figure}[!htbp]
\centering
\makebox[\columnwidth][l]{
   \hspace*{-0.7cm}
   \includegraphics[width=1.1\columnwidth]{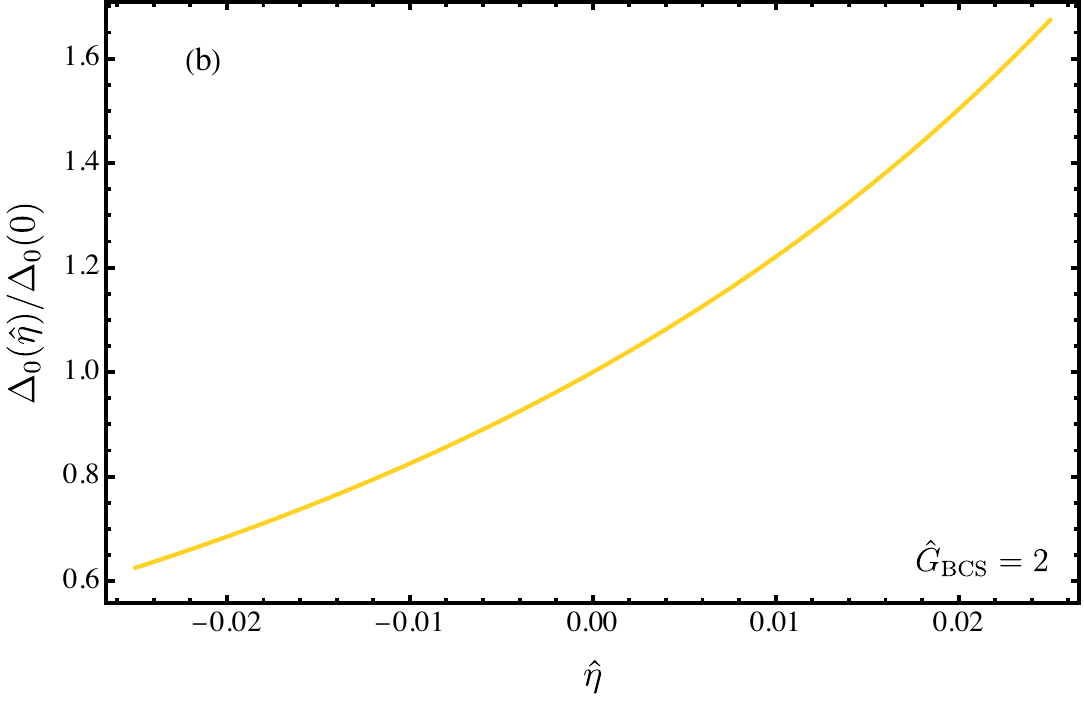}}
\caption{Zero temperature pairing gap normalized by its undeformed value, $\Delta_0(\widehat{\eta})/\Delta_0(0)$, for the same spectral parameters as in Fig.~\ref{fig:hl-density-of-states} and the coupling $\widehat{G}_{\mathrm{BCS}} = 2$.}
\label{fig:hl-pairing-gap}
\end{figure}

Fig.~\ref{fig:hl-pairing-gap} displays the corresponding change in the zero temperature gap.  The exponential dependence in Eq.~(\ref{eq:HL_BCS_weak_coupling_gap}) amplifies the moderate spectral variation seen in Fig.~\ref{fig:hl-density-of-states}: positive $\widehat{\eta}$ enhances the gap, while negative $\widehat{\eta}$ suppresses it. Notice that even within the effective regime, the higher spatial derivative term may produce a sizable change in the pairing scale without altering the weak coupling ratio $T_c/\Delta_0$. 

The factor in the exponent follows from the interaction normalization and the inclusion of both spin states in $N(0)$. The corresponding critical temperature satisfies
\begin{equation} 
T_c = \frac{e^{\gamma_{\mathrm{E}}}}{\pi}\Delta_0\simeq0.567\,\Delta_0, 
\label{eq:HL_BCS_ratio} 
\end{equation}
where $\gamma_{\mathrm{E}}$ is the Euler--Mascheroni constant. This ratio requires weak coupling, a nearly constant density of states, and an interaction without appreciable variation across the pairing shell. Within these assumptions, the modified dispersion changes the pairing scale through $p_F$ and $v_F$, while the exponential dependence on $N(0)$ can amplify a small spectral deformation.

To examine the electromagnetic response, we couple the fermions to a weak external field. The ordering of the spatial operators must be specified because covariant derivatives do not generally commute. With $D_i=\partial_i+ieA_i$, $\Pi_i=iD_i$, and $\boldsymbol{\Pi}^2 = \delta^{ij} \Pi_i \Pi_j$, we adopt the symmetrized prescription
\begin{equation} 
\mathcal{K}_{\mathrm{sp}}[A] = \frac{1}{2}\left\{\gamma^i\Pi_i,F\!\left(\sqrt{\boldsymbol{\Pi}^2}\right)\right\}, 
\label{eq:HL_BCS_gauge_covariant_operator}
\end{equation}
where the braces denote an anticommutator and the fractional power is defined spectrally. This prescription is gauge covariant, gives a Hermitian single-particle Hamiltonian, and reduces to $\gamma^ip_iF(p)$ at zero field.

For a uniform vector potential, the one-photon vertex is
\begin{equation} 
\Gamma_i(\boldsymbol{p}) = e\frac{\partial}{\partial p_i}\left[\gamma^kp_kF(p)\right]=e\left[\gamma^iF(p)+\gamma^kp_k\frac{p_i}{p}F'(p)\right]. 
\label{eq:HL_BCS_HL_vertex} 
\end{equation}
Here spatial momentum derivatives and contractions use $\delta^{ij}$. Substituting Eq.~\eqref{eq:HL_BCS_F_derivative} gives
\begin{equation} 
\Gamma_i(\boldsymbol{p}) = \frac{ep^{z-1}}{\Lambda_{\mathrm{HL}}^{z-1}}\left\{\gamma^i(1-\eta p^2)+\frac{\gamma^kp_kp_i}{p^2}\left[(z-1)-(z+1)\eta p^2\right]\right\}. 
\label{eq:HL_BCS_explicit_HL_vertex} 
\end{equation}
The two photon contact vertex is
\begin{equation} 
\Gamma_{ij}^{(2)}(\boldsymbol{p}) = e^2\frac{\partial^2}{\partial p_i\partial p_j}\left[\gamma^kp_kF(p)\right]. 
\label{eq:HL_BCS_contact_vertex} 
\end{equation}
It vanishes for a strictly linear spatial Dirac operator, but must be retained for the present deformation.

In the charge conjugate Nambu basis defined above, the two sectors carry opposite charges. Their zero momentum vertices are therefore
\begin{equation} 
\boldsymbol{\Gamma}_i(\boldsymbol{p}) = \begin{pmatrix}\Gamma_i(\boldsymbol{p})&0\\ 0&-\Gamma_i(\boldsymbol{p})\end{pmatrix}, \,\,\, \boldsymbol{\Gamma}_{ij}^{(2)}(\boldsymbol{p}) = \begin{pmatrix}\Gamma_{ij}^{(2)}(\boldsymbol{p})&0\\ 0&\Gamma_{ij}^{(2)}(\boldsymbol{p})
\end{pmatrix}. 
\label{eq:HL_BCS_Nambu_vertex}
\end{equation}
At finite momentum transfer, the vertices must be obtained by differentiating Eq.~\eqref{eq:HL_BCS_gauge_covariant_operator}; notice that the uniform field expressions alone do not determine them.

We fix the response convention through the quadratic change of the grand potential for a static Fourier mode,
\begin{equation} 
\delta\Omega = \frac{1}{2}A_i(-Q)K_{ij}(Q)A_j(Q), \quad Q = (0,\boldsymbol{q}), 
\label{eq:HL_BCS_response_convention} 
\end{equation}
so the induced current is $j_i(Q) = -K_{ij}(Q)A_j(Q)$. Define the vertices by the expansion $\mathcal{G}^{-1}[A] = \mathcal{G}^{-1}[0]-\boldsymbol{\Gamma}_i A_i+\boldsymbol{\Gamma}_{ij}^{(2)}A_iA_j/2+\cdots$, with momentum convolutions understood. Differentiating the fermionic contribution $-T\operatorname{Tr}\ln\mathcal{G}^{-1}/2$ twice at fixed $\Delta$ yields
\begin{widetext}
\begin{equation} 
K_{ij}^{(0)}(Q) = \frac{T}{2}\sum_n\int_{\boldsymbol{p}}\operatorname{Tr}\left[\mathcal{G}(P+Q)\boldsymbol{\Gamma}_i(P+Q,P)\mathcal{G}(P)\boldsymbol{\Gamma}_j(P,P+Q)-\mathcal{G}(P)\boldsymbol{\Gamma}_{ij}^{(2)}(P;Q,-Q)\right]. 
\label{eq:HL_BCS_response_kernel} 
\end{equation}
\end{widetext}
The trace includes Dirac and Nambu indices. The relative signs follow from the stated grand-potential convention. The ultraviolet regulator must also be implemented gauge covariantly, including the contributions generated by its field dependence; a cutoff imposed on canonical momentum without these terms can violate the Ward identity.

The physical response additionally includes the induced variation of the condensate. Writing $\Delta(x)=\Delta+\varphi_1(x)+i\varphi_2(x)$ and eliminating these amplitude and phase perturbations gives
\begin{equation} 
K_{ij}(Q)=K_{ij}^{(0)}(Q) - \mathcal{Q}_{ia}(Q)\left[\mathcal{M}(Q)^{-1}\right]_{ab}\mathcal{Q}_{bj}(Q), 
\label{eq:HL_BCS_collective_response} 
\end{equation}
where $\mathcal{M}_{ab}$ is the full order-parameter Hessian and $\mathcal{Q}_{ia}$ denotes the mixed electromagnetic--order-parameter derivatives of the grand potential. These quantities are evaluated at the self-consistent paired saddle. Treating the condensate response together with the fermionic vertices is required by gauge invariance \cite{Nambu:1960tm,GuoChienHe:2012}.

For an isotropic homogeneous state, the mixed static correlators are proportional to $q_i$, so the collective contribution is longitudinal. It restores the static Ward identity without changing the transverse screening coefficient. Decomposing
\begin{equation} 
K_{ij}(0,\boldsymbol{q}) = \left(\delta_{ij}-\frac{q_iq_j}{q^2}\right)K_T(0,\boldsymbol{q})+\frac{q_iq_j}{q^2}K_L(0,\boldsymbol{q}), 
\label{eq:HL_BCS_transverse_decomposition} 
\end{equation}
the Meissner screening mass is
\begin{equation} 
m_M^2 = \lim_{|\boldsymbol{q}|\rightarrow0}K_T(0,\boldsymbol{q}),
\label{eq:HL_BCS_Meissner_mass} 
\end{equation}
with the zero frequency limit taken first. A consistent calculation gives $m_M^2 = 0$ in the normal phase. For a canonically normalized Maxwell field in the local London limit,
\begin{equation} 
\lambda_L^{-2} = m_M^2. 
\label{eq:HL_BCS_London_depth} 
\end{equation}
A homogeneous superconducting phase requires a thermodynamically stable paired solution and $m_M^2>0$. The gap equation establishes pairing, while the transverse response determines magnetic screening. A negative stiffness signals an instability of the assumed homogeneous state.


\section{The effective interparticle potential}
\label{sec:HL_effective_potential}

The fermionic dispersion relation determines the poles of the free two point function, but an interparticle potential also requires an interaction channel and its coupling to the sources. Unlike the bosonic exchange considered in Ref.~\cite{AraujoFilho:2026yaj}, the kinetic deformation introduced above does not by itself specify such an interaction. The free fermion propagator is
\begin{equation} 
S_F(p_0,\boldsymbol{p}) = i\frac{\gamma^0p_0+\gamma^ip_iF(p)+m}{p_0^2-m^2-p^2F(p)^2+i0^+}, \qquad p=|\boldsymbol{p}|.
\label{eq:HL_fermion_propagator} 
\end{equation}

We introduce an effective spin independent exchange channel whose spectral denominator is chosen to coincide with the fermionic one. Its static kernel is therefore defined by
\begin{equation} 
\widetilde{G}_{\mathrm{st}}(p) \equiv \frac{1}{m^2+p^2F(p)^2}. 
\label{eq:HL_static_kernel_general} 
\end{equation}
This identification, including the use of the same mass parameter $m$, is an additional interaction assumption, not the static limit of the complete matrix valued propagator. We denote the signed exchange strength by $\mathfrak{g}=\sigma g_1g_2$, where $g_1$ and $g_2$ are the source couplings and $\sigma=\pm1$ accounts for the overall sign associated with the exchange channel. Spin  and velocity dependent contributions require explicit interaction vertices and are not included here.

In particular, for $z=2$, the spatial function becomes
\begin{equation}
F_2(p) = \frac{p}{\Lambda_{\mathrm{HL}}}\left(1-\eta p^2\right), \qquad \eta \equiv \frac{\bar{\alpha}}{M_P^2}. 
\label{eq:F_z2} 
\end{equation}
Taking $m>0$ and $\Lambda_{\mathrm{HL}}>0$, we introduce
\begin{equation} 
\varrho^4 \equiv m^2\Lambda_{\mathrm{HL}}^2, \qquad \varrho=\sqrt{m\Lambda_{\mathrm{HL}}}.
\label{eq:rho_definition} 
\end{equation}

The unexpanded kernel associated with the displayed kinetic truncation is
\begin{equation} 
\widetilde{G}_{\mathrm{st}}(p) = \frac{\Lambda_{\mathrm{HL}}^2}{\varrho^4+p^4(1-\eta p^2)^2}. 
\label{eq:HL_static_kernel_unexpanded} 
\end{equation}

Expanding to first order in $\eta$ gives
\begin{equation} 
\widetilde{G}_{\mathrm{st}}(p) = \frac{\Lambda_{\mathrm{HL}}^2}{p^4+\varrho^4}\left[1+\frac{2\eta p^6}{p^4+\varrho^4}\right]+\mathcal{O}(\eta^2). 
\label{eq:HL_static_kernel_z2} 
\end{equation}
For $m>0$, the denominator in Eq.~\eqref{eq:HL_static_kernel_unexpanded} has no zeros at real momentum. Its additional complex roots at momenta of order $|\eta|^{-1/2}$ lie outside the derivative expansion and cannot be treated as controlled predictions. To resolve the screening profile within the effective theory, we choose a momentum cutoff $\Lambda_{\mathrm{cut}}$ below its breakdown scale such that
\begin{equation} 
\varrho \ll \Lambda_{\mathrm{cut}}, \qquad |\eta|\Lambda_{\mathrm{cut}}^2 \ll 1.
\label{eq:HL_potential_hierarchy} 
\end{equation}

The continuum Fourier transforms below represent the resolved exchange contribution at distances $r\gg\Lambda_{\mathrm{cut}}^{-1}$, assuming no additional light exchange modes. Ultraviolet sensitive contact and short range contributions require matching. Extending the expanded integrand to infinite momentum is a Fourier prescription.

In this manner, the potential associated with the displayed kernel is
\begin{equation} 
V(r) = \mathfrak{g}\int\frac{\mathrm{d}^3p}{(2\pi)^3}\,e^{i\boldsymbol{p}\cdot\boldsymbol{r}}\widetilde{G}_{\mathrm{st}}(p), \qquad r=|\boldsymbol{r}|. 
\label{eq:HL_potential_Fourier} 
\end{equation}

For an isotropic function $\widetilde{H}(p)$, angular integration yields \cite{AraujoFilho:2025fwd}
\begin{equation} 
\int\frac{\mathrm{d}^3p}{(2\pi)^3}\,e^{i\boldsymbol{p}\cdot\boldsymbol{r}}\widetilde{H}(p) = \frac{1}{2\pi^2r}\int_0^\infty\mathrm{d}p\,p\sin(pr)\widetilde{H}(p). 
\label{eq:HL_radial_Fourier_formula} 
\end{equation}

The zeroth order denominator factorizes as $p^4+\varrho^4 = (p^2+i\varrho^2)(p^2-i\varrho^2)$. Its upper half plane poles are $p = \varrho e^{i\pi/4}$ and $p=\varrho e^{3i\pi/4}$, and the residue calculation gives, for $r>0$,
\begin{equation} 
\int\frac{\mathrm{d}^3p}{(2\pi)^3}\,\frac{e^{i\boldsymbol{p}\cdot\boldsymbol{r}}}{p^4+\varrho^4} = \frac{e^{-x}\sin x}{4\pi\varrho^2r}, \qquad x\equiv\frac{\varrho r}{\sqrt{2}}. 
\label{eq:transform_biharmonic_massive} 
\end{equation}

The first order transform can be obtained by differentiating the auxiliary integral
\begin{equation} 
J(r;\varrho) \equiv \int\frac{\mathrm{d}^3p}{(2\pi)^3}\,\frac{p^2e^{i\boldsymbol{p}\cdot\boldsymbol{r}}}{p^4+\varrho^4} = \frac{e^{-x}\cos x}{4\pi r}. 
\label{eq:HL_auxiliary_transform} 
\end{equation}

Indeed, the identity
\begin{equation} 
\frac{p^6}{(p^4+\varrho^4)^2} = \left(1+\frac{\varrho}{4}\frac{\partial}{\partial\varrho}\right)\frac{p^2}{p^4+\varrho^4} 
\label{eq:HL_correction_identity} 
\end{equation}
implies, with the derivative taken at fixed $r$,
\begin{equation} 
\int\frac{\mathrm{d}^3p}{(2\pi)^3}\,\frac{p^6e^{i\boldsymbol{p}\cdot\boldsymbol{r}}}{(p^4+\varrho^4)^2} = \frac{e^{-x}}{4\pi r}\left[\cos x-\frac{x}{4}\left(\cos x+\sin x\right)\right]. 
\label{eq:transform_HL_correction} 
\end{equation}

Substitution into Eq.~\eqref{eq:HL_potential_Fourier} yields
\begin{equation} 
V(r) = \frac{\mathfrak{g}\Lambda_{\mathrm{HL}}^2e^{-x}}{4\pi r}\left\{\frac{\sin x}{\varrho^2}+2\eta\left[\cos x-\frac{x}{4}\left(\cos x+\sin x\right)\right]\right\}. 
\label{eq:HL_effective_potential} 
\end{equation}
At $\eta=0$, the quartic spatial operator produces an exponentially damped oscillatory potential with decay length $\sqrt{2}/\varrho$. It is important to notice that reversing the sign of $\mathfrak{g}$ reverses the potential, but attraction and repulsion are determined locally by the radial force $-\mathrm{d}V/\mathrm{d}r$, not by the sign of $V$ alone.

\begin{figure}[!htbp]
\centering
\makebox[\columnwidth][l]{
   \hspace*{-0.7cm}
   \includegraphics[width=1.1\columnwidth]{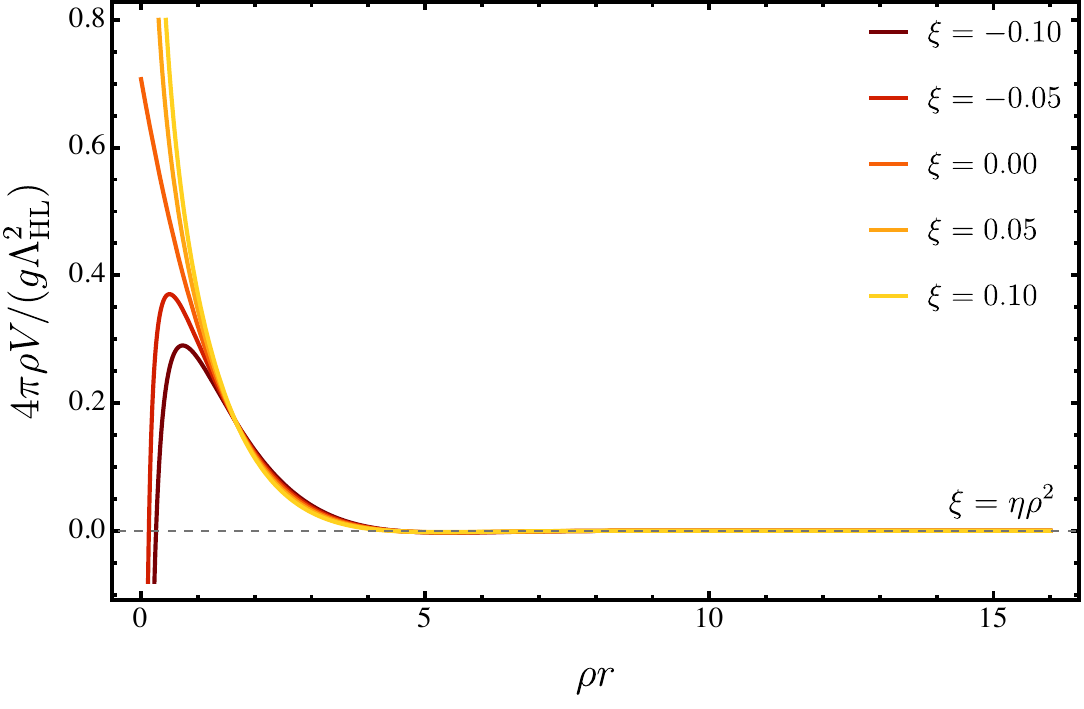}}
\caption{Dimensionless massive potential, $4\pi\rho V/(g\Lambda_{\mathrm{HL}}^2)$, as a function of $\rho r$ for $\xi\equiv\eta\rho^2=\{-0.10,-0.05,0,0.05,0.10\}$.  The horizontal dashed line marks the zeros of the oscillatory profile.}
\label{fig:hl-massive-potential}
\end{figure}

The radial structure of the exchange interaction is exhibited in Fig.~\ref{fig:hl-massive-potential}.  Every curve shows the damped oscillatory profile generated by the quartic denominator, while the deformation changes the short distance amplitude and the first oscillation.  Its influence rapidly decreases for $\rho r\geq 5$, where the exponential damping controls the profile. 

The terms proportional to $x$ encode changes in both the damping rate and the oscillation wavenumber. Perturbing the two low momentum poles in the upper half plane gives
\begin{equation} 
p_\pm = \frac{\varrho}{\sqrt{2}}\left[\pm\left(1-\frac{\eta\varrho^2}{2}\right)+i\left(1+\frac{\eta\varrho^2}{2}\right)\right]+\mathcal{O}(\eta^2\varrho^5).
\label{eq:HL_shifted_static_poles} 
\end{equation}
Positive $\eta$ increases the damping rate and decreases the oscillation wavenumber. Eq.~\eqref{eq:HL_effective_potential} is an expansion at fixed $r$; the accumulated phase and damping corrections remain perturbative when $|\eta|\varrho^2x\ll1$. Also, it is worthy to be commented that a uniform approximation at larger distances requires retaining the shifted poles in the exponential factors.

Expanding the coefficient functions for $x\ll1$, we obtain
\begin{equation} 
V(r) = \frac{\mathfrak{g}\Lambda_{\mathrm{HL}}^2}{4\pi}\left[\frac{2\eta}{r}+\frac{1-\frac{5}{2}\eta\varrho^2}{\sqrt{2}\varrho}-\frac{r}{2}+\mathcal{O}(\varrho r^2,\eta\varrho^3r^2)\right]  +  \mathcal{O}(\eta^2). 
\label{eq:HL_short_distance_potential} 
\end{equation}
The zeroth order continuum kernel is finite at the origin, whereas its first order correction contains a $1/r$ term. Within the effective theory, this expansion describes the intermediate window $\Lambda_{\mathrm{cut}}^{-1}\ll r\ll\varrho^{-1}$, not the limit $r\to0$. Notice that neither the finite zeroth order value nor the apparent singularity determines the ultraviolet behavior of the physical interaction.

On the other hand, the massless limit requires an infrared subtraction. Setting $m=0$ directly in the momentum space kernel gives
\begin{equation} 
\widetilde{G}_{\mathrm{st}}^{(m=0)}(p) = \Lambda_{\mathrm{HL}}^2\left(\frac{1}{p^4}+\frac{2\eta}{p^2}\right)+\mathcal{O}(\eta^2). 
\label{eq:HL_massless_static_kernel} 
\end{equation}

The transform of $1/p^4$ contains a distance independent infrared divergence. Removing this constant, the relevant real Fourier transforms are
\begin{equation} 
\int\frac{\mathrm{d}^3p}{(2\pi)^3}\,\frac{\cos(\boldsymbol{p}\cdot\boldsymbol{r})-1}{p^4} = -\frac{r}{8\pi}, \qquad \int\frac{\mathrm{d}^3p}{(2\pi)^3}\,\frac{e^{i\boldsymbol{p}\cdot\boldsymbol{r}}}{p^2} = \frac{1}{4\pi r}. 
\label{eq:HL_massless_transforms} 
\end{equation}
The massless potential is 
\begin{equation} 
V_{m=0}(r) = \mathfrak{g}\Lambda_{\mathrm{HL}}^2\left(-\frac{r}{8\pi}+\frac{\eta}{2\pi r}\right)+\mathcal{O}(\eta^2), 
\label{eq:HL_massless_potential} 
\end{equation}
up to an additive constant. The same expression follows from Eq.~\eqref{eq:HL_effective_potential} by subtracting $\mathfrak{g}\Lambda_{\mathrm{HL}}^2/(4\pi\sqrt{2}\varrho)$ before taking $\varrho\to0$ at fixed $r$.

Within the pure quartic model, the magnitude of the leading massless term grows linearly with distance. Its radial force is constant and repulsive for $\mathfrak{g}>0$, becoming attractive when the sign is reversed. Since the potential does not approach a finite constant at spatial infinity, the usual free plane wave scattering boundary conditions do not apply. We, in this case, use the massive kernel in the scattering analysis. Its exponential decay permits the standard short range formulation, while the first Born approximation additionally requires weak scattering, as in the treatment of screened interactions in Ref.~\cite{Touati:2025}.


\section{Electron scattering by a static source}
\label{sec:HL_electron_scattering}

We investigate elastic electron scattering by the static potential obtained in the preceding section, neglecting recoil and internal excitations of the source. The calculation concerns an external potential, not the antisymmetrized amplitude for a two-electron collision. We adopt natural units, $\hbar=c=1$, and denote the incident and outgoing momenta by $\boldsymbol{k}_i$ and $\boldsymbol{k}_f$, with $|\boldsymbol{k}_i|=|\boldsymbol{k}_f|=\kappa>0$. The momentum transfer is
\begin{equation} 
\boldsymbol{q}=\boldsymbol{k}_f-\boldsymbol{k}_i, \qquad q^2 = 2\kappa^2(1-\cos\theta) = 4\kappa^2\sin^2\left(\frac{\theta}{2}\right). 
\label{eq:HL_momentum_transfer_intermediate} 
\end{equation}

We retain the identification $m=m_e>0$ in the exchange ansatz, so that $\varrho^4=m_e^2\Lambda_{\mathrm{HL}}^2$. Its momentum-space potential is
\begin{equation} 
\widetilde{V}(q) = \frac{\mathfrak{g}\Lambda_{\mathrm{HL}}^2}{q^4+\varrho^4}\left[1+\frac{2\eta q^6}{q^4+\varrho^4}\right]+\mathcal{O}(\eta^2), \qquad \eta=\frac{\bar{\alpha}}{M_P^2}. 
\label{eq:HL_momentum_potential} 
\end{equation}

All results below refer to this exchange contribution, without additional contact interactions. Since $0\leq q\leq2\kappa$, a calculation over the full angular range requires $2\kappa\ll\Lambda_{\mathrm{cut}}$ and $|\eta|(2\kappa)^2\ll1$, together with the hierarchy specified in the preceding section.

To fix the electron vertex, we choose the density interaction $\mathcal{H}_{\mathrm{int}}=V(r)\psi^\dagger\psi=V(r)\bar{\psi}\gamma^0\psi$. The one-particle Hamiltonian, restricted to the effective momentum domain, is
\begin{equation} 
\begin{split}
& H = H_0(-i\boldsymbol{\nabla}) + V(r)\mathbf{1}_4, \,\,\,  H_0(\boldsymbol{p}) = \boldsymbol{\alpha}\cdot\boldsymbol{p}\,F_2(p) + \beta m_e,
\label{eq:HL_scattering_Hamiltonian} 
\end{split}
\end{equation}
with $\alpha^i = \gamma^0\gamma^i$, and $ \beta = \gamma^0$.

For the displayed quadratic truncation, the positive electron energy and its group velocity are
\begin{equation} 
E_\kappa=\sqrt{m_e^2+\frac{\kappa^4}{\Lambda_{\mathrm{HL}}^2}(1-\eta\kappa^2)^2}, 
\label{eq:HL_electron_energy} 
\end{equation}
and
\begin{equation} v_g(\kappa)=\frac{\mathrm{d}E_\kappa}{\mathrm{d}\kappa}=\frac{2\kappa^3}{\Lambda_{\mathrm{HL}}^2E_\kappa}(1-\eta\kappa^2)(1-2\eta\kappa^2). \label{eq:HL_group_velocity} \end{equation}
Within the effective regime, $v_g(\kappa)>0$ and the elastic energy shell has a unique momentum magnitude. The unexpanded expressions organize the calculation; terms beyond first order in $\eta$ are not retained.

Let $u_s(\boldsymbol{p})$ be positive-energy eigenspinors normalized by $u_s^\dagger(\boldsymbol{p})u_{s'}(\boldsymbol{p})=\delta_{ss'}$. Their completeness relation is
\begin{equation} 
P_+(\boldsymbol{p})\equiv\sum_su_s(\boldsymbol{p})u_s^\dagger(\boldsymbol{p}) = \frac{1}{2}\left[\mathbf{1}_4+\frac{H_0(\boldsymbol{p})}{E(p)}\right]. 
\label{eq:HL_positive_energy_projector} 
\end{equation}

The scattering spinor satisfies the Lippmann--Schwinger equation~\cite{LippmannSchwinger:1950},
\begin{equation} 
\psi_{\boldsymbol{k}_i,s_i}^{(+)}(\boldsymbol{r}) = u_{s_i}(\boldsymbol{k}_i)e^{i\boldsymbol{k}_i\cdot\boldsymbol{r}} + \int\mathrm{d}^3r'\,\mathcal{G}_0^{(+)}(\boldsymbol{r}-\boldsymbol{r}';E_\kappa)V(r')\psi_{\boldsymbol{k}_i,s_i}^{(+)}(\boldsymbol{r}'), 
\label{eq:HL_Lippmann_Schwinger} 
\end{equation}
where the matrix valued outgoing Green function is
\begin{equation} 
\mathcal{G}_0^{(+)}(\boldsymbol{R};E_\kappa) = \int_{|\boldsymbol{p}|<\Lambda_{\mathrm{cut}}}\frac{\mathrm{d}^3p}{(2\pi)^3}\,e^{i\boldsymbol{p}\cdot\boldsymbol{R}}\left[E_\kappa-H_0(\boldsymbol{p})+i0^+\right]^{-1}.
\label{eq:HL_free_Green_momentum} 
\end{equation}

The cutoff excludes momenta outside the effective description without changing the elastic pole. Its positive-energy residue is $P_+(\boldsymbol{p})$, and near $p=\kappa$,
\begin{equation} 
E_\kappa-E(p) = -v_g(\kappa)(p-\kappa)+\mathcal{O}\bigl((p-\kappa)^2\bigr). 
\label{eq:HL_pole_expansion} 
\end{equation}

Evaluating the outgoing contribution at large $R=|\boldsymbol{R}|$ gives
\begin{equation} 
\mathcal{G}_0^{(+)}(\boldsymbol{R};E_\kappa)\simeq-\frac{\kappa}{2\pi v_g(\kappa)}\frac{e^{i\kappa R}}{R}P_+(\kappa\widehat{\boldsymbol{R}}). 
\label{eq:HL_asymptotic_Green} 
\end{equation}

The asymptotic state can thus be written as
\begin{equation} 
\psi_{\boldsymbol{k}_i,s_i}^{(+)}(\boldsymbol{r})\simeq u_{s_i}(\boldsymbol{k}_i)e^{i\boldsymbol{k}_i\cdot\boldsymbol{r}}+\frac{e^{i\kappa r}}{r}\sum_{s_f}f_{s_fs_i}(\theta,\phi)u_{s_f}(\boldsymbol{k}_f),
\label{eq:HL_asymptotic_wavefunction} 
\end{equation}
with $\boldsymbol{k}_f = \kappa\widehat{\boldsymbol{r}}$.

Replacing the spinor under the integral in Eq.~\eqref{eq:HL_Lippmann_Schwinger} by the incident state gives the first Born amplitude~\cite{Born:1926},
\begin{equation} 
f_{s_fs_i}^{\mathrm{B}}(\theta,\phi) = f_{\mathrm{B}}(\theta)u_{s_f}^\dagger(\boldsymbol{k}_f)u_{s_i}(\boldsymbol{k}_i), \,\,\, f_{\mathrm{B}}(\theta) = - \frac{\kappa}{2\pi v_g(\kappa)}\widetilde{V}(q). 
\label{eq:HL_Born_general} 
\end{equation}
Here $f_{\mathrm{B}}$ is the orbital coefficient, not the complete spin-resolved amplitude. For ordinary nonrelativistic electron kinematics, $v_g=\kappa/m_e$, it reduces to
\begin{equation} 
f_{\mathrm{B}}(\theta) = -\frac{m_e}{2\pi}\widetilde{V}(q) = -\frac{2m_e}{q}\int_0^\infty\mathrm{d}r\,rV(r)\sin(qr), 
\label{eq:HL_Born_nonrelativistic} 
\end{equation}
as in the potential scattering prescription of Refs.~\cite{Touati:2025,AraujoFilho:2026yaj,AraujoFilho:2026oqc}. This replacement is not the low momentum limit of the pure $z=2$ dispersion, whose kinetic energy is quartic.

The probability current of a normalized free electron is $v_g(\kappa)\widehat{\boldsymbol{k}}$. Since the incoming and outgoing speeds are equal, the unpolarized differential cross section is
\begin{equation} 
\frac{\mathrm{d}\sigma}{\mathrm{d}\Omega} = \frac{1}{2}\sum_{s_i,s_f}\left|f_{s_fs_i}^{\mathrm{B}}\right|^2=\mathcal{M}_{\mathrm{HL}}(\theta)\left|f_{\mathrm{B}}(\theta)\right|^2. 
\label{eq:HL_Mott_cross_section} 
\end{equation}

The spin sum follows directly from the projectors:
\begin{equation} 
\mathcal{M}_{\mathrm{HL}}(\theta)=\frac{1}{2}\operatorname{tr}\!\left[P_+(\boldsymbol{k}_f)P_+(\boldsymbol{k}_i)\right]=1-\beta_{\mathrm{HL}}^2\sin^2\left(\frac{\theta}{2}\right),
\label{eq:HL_Mott_factor} 
\end{equation}
where $\beta_{\mathrm{HL}} = \kappa F_2(\kappa)/E_\kappa$.

This is the counterpart of the Mott factor for the adopted fermionic operator~\cite{Mott:1929}. The quantity $\beta_{\mathrm{HL}}$ measures the spinor mixing and is not the group velocity. They coincide for $F(p)=1$, but not for the Lifshitz dispersion.

Using Eq.~\eqref{eq:HL_momentum_potential}, we obtain
\begin{equation} 
\begin{split}
&\frac{\mathrm{d}\sigma}{\mathrm{d}\Omega} = \frac{\kappa^2\mathfrak{g}^{2}\Lambda_{\mathrm{HL}}^4}{4\pi^2v_g(\kappa)^2(q^4+\varrho^4)^2}\left[1+\frac{4\eta q^6}{q^4+\varrho^4}\right] \left[1-\beta_{\mathrm{HL}}^2\sin^2\left(\frac{\theta}{2}\right)\right]. 
\label{eq:HL_differential_cross_section} 
\end{split}
\end{equation}

The deformation enters through the group velocity, the exchange kernel, and the external spinors. To display all three contributions at fixed incident momentum, we define
\begin{equation} 
E_0 = \sqrt{m_e^2+\frac{\kappa^4}{\Lambda_{\mathrm{HL}}^2}}, \qquad v_0=\frac{2\kappa^3}{\Lambda_{\mathrm{HL}}^2E_0}, \qquad \beta_0^2=\frac{\kappa^4}{\Lambda_{\mathrm{HL}}^2E_0^2}. 
\label{eq:HL_zeroth_order_kinematics} 
\end{equation}

The required expansions are
\begin{equation} 
E_\kappa = E_0 - \eta\frac{\kappa^6}{\Lambda_{\mathrm{HL}}^2E_0}+\mathcal{O}
(\eta^2), \label{eq:HL_energy_expansion} 
\end{equation}
\begin{equation} 
v_g(\kappa) = v_0\left[1-\eta\kappa^2(3-\beta_0^2)\right]+\mathcal{O}(\eta^2), 
\label{eq:HL_velocity_expansion} 
\end{equation}
and
\begin{equation} \beta_{\mathrm{HL}}^2=\beta_0^2\left[1-2\eta\kappa^2(1-\beta_0^2)\right]+\mathcal{O}(\eta^2). \label{eq:HL_spinor_ratio_expansion} \end{equation}
Writing $\mathcal{M}_0(\theta)=1-\beta_0^2\sin^2(\theta/2)$, the differential cross section becomes
\begin{widetext}
\begin{equation} 
\frac{\mathrm{d}\sigma}{\mathrm{d}\Omega} = \frac{\mathfrak{g}^{2}\Lambda_{\mathrm{HL}}^8E_0^2}{16\pi^2\kappa^4(q^4+\varrho^4)^2}\left\{\mathcal{M}_0(\theta)\left[1+\eta\left(2\kappa^2(3-\beta_0^2)+\frac{4q^6}{q^4+\varrho^4}\right)\right]+2\eta\kappa^2\beta_0^2(1-\beta_0^2)\sin^2\left(\frac{\theta}{2}\right)\right\}+\mathcal{O}(\eta^2).
\label{eq:HL_differential_cross_section_expanded}
\end{equation}
\end{widetext}
The term proportional to $\beta_0^2(1-\beta_0^2)$ accounts for the deformation of the external spinors. At fixed $\kappa$, positive $\eta$ enhances the differential cross section within the perturbative regime.

At fixed $\kappa>0$, the massive kernel removes the forward singularity. Since $\mathcal{M}_{\mathrm{HL}}(0)=1$,
\begin{equation} 
\left.\frac{\mathrm{d}\sigma}{\mathrm{d}\Omega}\right|_{\theta\to0} = \frac{\kappa^2\mathfrak{g}^{2}\Lambda_{\mathrm{HL}}^4}{4\pi^2v_g(\kappa)^2\varrho^8}+\mathcal{O}
(\eta^2). 
\label{eq:HL_forward_cross_section} 
\end{equation}

In the window $\varrho\ll q\leq2\kappa\ll\Lambda_{\mathrm{cut}}$, the zeroth order angular dependence is
\begin{equation} 
\left(\frac{\mathrm{d}\sigma}{\mathrm{d}\Omega}\right)_{\eta = 0}\propto\frac{\mathcal{M}_0(\theta)}{q^8}. 
\label{eq:HL_high_q_scaling} 
\end{equation}
The $q^{-8}$ dependence belongs to the orbital factor, compared with $q^{-4}$ for Coulomb exchange; the electron spin factor remains angle dependent. In this window, the relative correction from the exchange kernel approaches $4\eta q^2$.

\begin{figure}[!htbp]
\centering
\makebox[\columnwidth][l]{
   \hspace*{-1cm}
   \includegraphics[width=1.1\columnwidth]{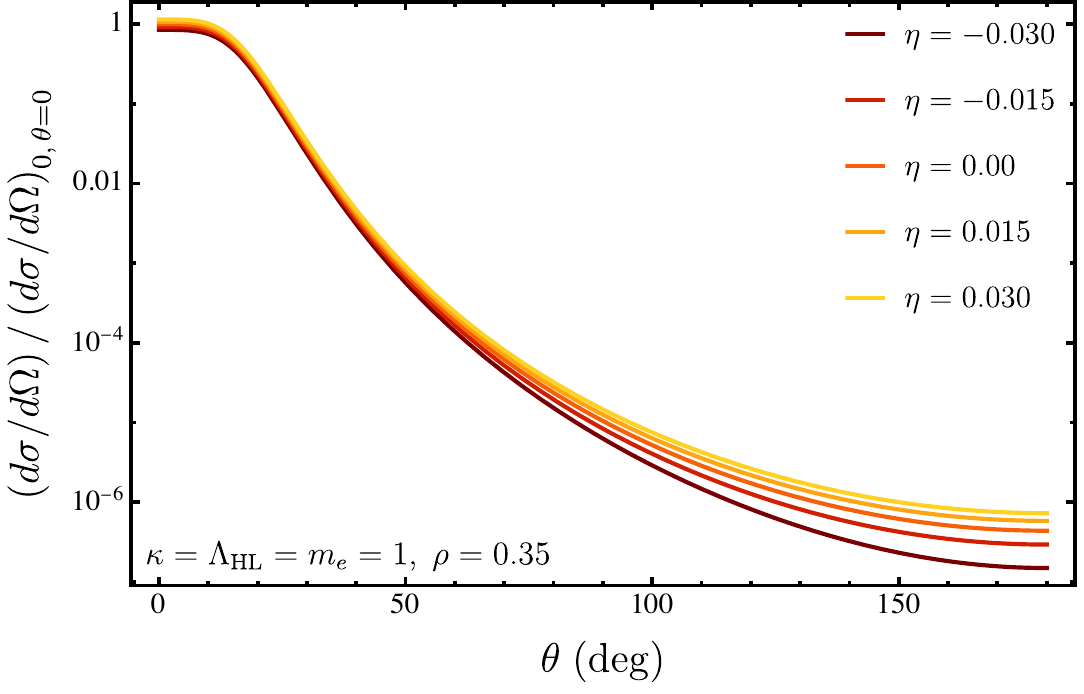}}
\caption{Unpolarized Born differential cross section, normalized by the undeformed forward value, as a function of the scattering angle for $\eta=\{-0.030,-0.015,0,0.015,0.030\}$, with $\kappa=\Lambda_{\mathrm{HL}}=m_e=1$ and $\rho=0.35$.  The vertical scale is logarithmic.}
\label{fig:hl-differential-cross-section}
\end{figure}

Fig.~\ref{fig:hl-differential-cross-section} shows that the massive kernel removes the forward singularity and produces a finite value at $\theta = 0$. Beyond the screening dominated region, the cross section falls rapidly with angle, reflecting the $q^{-8}$ orbital behavior identified in Eq.~(\ref{eq:HL_high_q_scaling}). Positive $\eta$ increases the cross section at fixed incident momentum, and the separation among the curves becomes more visible in the large angle tail, where the exchange kernel, group velocity, and spinor overlap corrections act together.

For the total cross section, we use $Q=2\kappa$ and $\sin\theta\,\mathrm{d}\theta=q\,\mathrm{d}q/\kappa^2$. Angular integration gives
\begin{equation} 
\sigma_{\mathrm{tot}}=\frac{\mathfrak{g}^{2}\Lambda_{\mathrm{HL}}^4}{2\pi v_g(\kappa)^2}\int_0^Q\frac{q\,\mathrm{d}q}{(q^4+\varrho^4)^2}\left(1-\frac{\beta_{\mathrm{HL}}^2q^2}{4\kappa^2}\right)\left(1+\frac{4\eta q^6}{q^4+\varrho^4}\right)\!.
\label{eq:HL_total_cross_section_integral} 
\end{equation}

The four required integrals are
\begin{equation} 
\mathcal{I}_0(Q)\equiv\int_0^Q\frac{q\,\mathrm{d}q}{(q^4+\varrho^4)^2}=\frac{Q^2}{4\varrho^4(Q^4+\varrho^4)}+\frac{1}{4\varrho^6}\tan^{-1}\left(\frac{Q^2}{\varrho^2}\right), 
\label{eq:HL_I0} 
\end{equation}
\begin{equation} 
\mathcal{I}_1(Q)\equiv\int_0^Q\frac{q^7\,\mathrm{d}q}{(q^4+\varrho^4)^3} = \frac{Q^8}{8\varrho^4(Q^4+\varrho^4)^2}, 
\label{eq:HL_I1} 
\end{equation}
\begin{equation} 
\mathcal{J}_0(Q)\equiv\int_0^Q\frac{q^3\,\mathrm{d}q}{(q^4+\varrho^4)^2}=\frac{Q^4}{4\varrho^4(Q^4+\varrho^4)}, 
\label{eq:HL_J0} 
\end{equation}
and
\begin{equation} 
\begin{split}
\mathcal{J}_1(Q) \equiv &\int_0^Q\frac{q^9\,\mathrm{d}q}{(q^4+\varrho^4)^3} = \frac{3}{16\varrho^2}\tan^{-1}\left(\frac{Q^2}{\varrho^2}\right) \\
& - \frac{Q^2(5Q^4+3\varrho^4)}{16(Q^4+\varrho^4)^2}. 
\label{eq:HL_J1} 
\end{split}
\end{equation}

The density coupled total cross section is 
\begin{equation} 
\begin{split}
\sigma_{\mathrm{tot}} = &\frac{\mathfrak{g}^{2}\Lambda_{\mathrm{HL}}^4}{2\pi v_g(\kappa)^2}\left[\mathcal{I}_0(Q) - \frac{\beta_{\mathrm{HL}}^2}{4\kappa^2}\mathcal{J}_0(Q) \right. \\
& \left. + 4\eta\left(\mathcal{I}_1(Q) -\frac{\beta_{\mathrm{HL}}^2}{4\kappa^2}\mathcal{J}_1(Q)\right)\right]. 
\label{eq:HL_total_cross_section} 
\end{split}
\end{equation}
Both $v_g$ and $\beta_{\mathrm{HL}}$ in this expression must be expanded through first order in $\eta$. Omitting the $\mathcal{J}_0$ and $\mathcal{J}_1$ terms gives the spinless approximation, valid when $\kappa F_2(\kappa)\ll m_e$. For $\varrho>0$ and fixed $\kappa>0$, the full angular integral is finite without an angular cutoff, unlike the Coulombic interaction considered in Ref.~\cite{AraujoFilho:2026yaj}.

Finiteness of the angular integral does not ensure the validity of the Born approximation near the threshold. For $\kappa\ll\varrho$, the spinor factor approaches unity, $v_g\simeq2\kappa^3/(m_e\Lambda_{\mathrm{HL}}^2)$, and the leading Born expressions are
\begin{equation}
f_{\mathrm{B}}\simeq-\frac{\mathfrak{g}\Lambda_{\mathrm{HL}}^2}{4\pi m_e\kappa^2}, \qquad \sigma_{\mathrm{tot}}^{\mathrm{B}}\simeq\frac{\mathfrak{g}^{2}\Lambda_{\mathrm{HL}}^4}{4\pi m_e^2\kappa^4}. 
\label{eq:HL_threshold_Born} 
\end{equation}
The dominant $s$--wave must remain perturbative, $|\kappa f_{\mathrm{B}}|\ll1$, which requires $|\mathfrak{g}|\Lambda_{\mathrm{HL}}^2/4\pi m_e\kappa \ll 1$. 
This necessary condition fails at sufficiently small $\kappa$ for any fixed nonzero coupling. The apparent $\kappa^{-4}$ growth cannot be extrapolated into that regime, where the first Born approximation must be replaced by a nonperturbative treatment.

A scalar bilinear represents a different interaction. If the Hamiltonian coupling is instead $H_{\mathrm{int}}=\beta V(r)$, the same orbital coefficient multiplies $u_{s_f}^\dagger\beta u_{s_i}$, and its spin average is
\begin{equation} 
\mathcal{M}_{\mathrm{S}}(\theta)=\frac{1}{2}\operatorname{tr}\!\left[P_+(\boldsymbol{k}_f)\beta P_+(\boldsymbol{k}_i)\beta\right] = 1-\beta_{\mathrm{HL}}^2\cos^2\left(\frac{\theta}
{2}\right). \label{eq:HL_scalar_spin_factor} 
\end{equation}
For that choice, $\mathcal{M}_{\mathrm{HL}}$ must be replaced by $\mathcal{M}_{\mathrm{S}}$ before angular integration.


\section{Atomic spectroscopy bounds}
\label{sec:atomic-bounds}

We treat the static interaction as a weak, spin independent addition to the ordinary Coulomb Hamiltonian, 
\begin{equation} 
H_{\mathrm{C}} = \frac{\mathbf{p}^{2}}{2\mu_{r}} - \frac{Z\alpha_{\mathrm{em}}}{r}, \qquad \mu_{r} = \frac{m_{\ell}M_{N}}{m_{\ell}+M_{N}}, \qquad a = \frac{1}{Z\alpha_{\mathrm{em}}\mu_{r}}. 
\label{eq:bohr-radius} 
\end{equation} 
This description assumes an infrared completion with a conventional quadratic kinetic energy; the pure $z=2$ dispersion does not admit the hydrogenic wave functions used below. The resulting bounds isolate the additional static interaction and require any residual modification of the atomic kinetic energy to be negligible at the stated accuracy. The smaller Bohr radius of muonic atoms enhances their sensitivity to short-range interactions, although nuclear-structure effects also become more important \cite{Pachucki2024}.

For a pointlike nucleus, the potential reads 
\begin{widetext} 
\begin{equation} 
V_{\mathrm{HL}}(r) = \mathcal{C}\frac{e^{-\omega r}}{r}\left\{\frac{\sin(\omega r)}{\varrho^{2}} + 2\eta\cos(\omega r) - \frac{\eta\omega r}{2}\left[\cos(\omega r)+\sin(\omega r)\right]\right\} + \mathcal{O}(\eta^{2}), \quad \mathcal{C} = \frac{\mathfrak{g}_{N\ell}\Lambda_{\mathrm{HL}}^{2}}{4\pi}, \quad \omega = \frac{\varrho}{\sqrt{2}}, \quad \varrho^{4} = m^{2}\Lambda_{\mathrm{HL}}^{2}. 
\label{eq:VHL-atomic} 
\end{equation} 
\end{widetext} 
Here $\mathfrak{g}_{N\ell}\equiv\sigma g_{N}g_{\ell}$ is the signed coupling product, with $\sigma=\pm1$ fixed by the exchange channel, and $\eta=\bar{\alpha}/M_{P}^{2}$.

The derivative expansion must remain controlled over the momentum scales relevant to the matrix elements. Introducing the effective theory cutoff $\Lambda_{\mathrm{cut}}$, we require \begin{equation} 
p_{B}\equiv a^{-1}, \qquad \max(p_{B},\varrho)\ll\Lambda_{\mathrm{cut}}, \qquad |\eta|\Lambda_{\mathrm{cut}}^{2}\ll1. 
\label{eq:atomic-EFT-regime} 
\end{equation} 
The radial integrals below employ the continuum form of the truncated potential. Their interpretation therefore assumes that independent short distance lepton--nucleus operators have been fixed by matching or are negligible; otherwise, their coefficients must enter the spectroscopic analysis.

At first order in the additional interaction, the level displacement is 
\begin{equation} 
\delta E_{nL} = \langle nLM_{L}|V_{\mathrm{HL}}|nLM_{L}\rangle,
\label{eq:first-order-shift} 
\end{equation} 
which is independent of $M_{L}$ because the perturbation is central. This approximation requires state mixing and higher-order energy shifts to remain negligible, including near cancellations in a transition shift. To evaluate the radial integrals, we define 
\begin{equation} 
\mathcal{M}_{j}(\beta) = \int_{0}^{\infty}\mathrm{d}r\,r^{j}e^{-\beta r}V_{\mathrm{HL}}(r), \qquad \zeta_{\beta} = \beta+\omega-i\omega, 
\label{eq:moment-definition} 
\end{equation} 
whrere $\beta>0$, and $j\geq2$. Direct integration gives 
\begin{widetext} 
\begin{equation} 
\mathcal{M}_{j}(\beta) = \mathcal{C}\Gamma(j)\left\{\frac{\operatorname{Im}(\zeta_{\beta}^{-j})}{\varrho^{2}} + 2\eta\operatorname{Re}(\zeta_{\beta}^{-j}) - \frac{\eta\omega j}{2}\left[\operatorname{Re}(\zeta_{\beta}^{-j-1})+\operatorname{Im}(\zeta_{\beta}^{-j-1})\right]\right\} + \mathcal{O}(\eta^{2}). 
\label{eq:closed-moment} 
\end{equation} 
\end{widetext}

The normalized hydrogenic wave functions then yield 
\begin{equation} 
\delta E_{1S} = \frac{4}{a^{3}}\mathcal{M}_{2}\left(\frac{2}{a}\right), 
\label{eq:shift-1S} 
\end{equation} 
\begin{equation} 
\delta E_{2S} = \frac{1}{8a^{3}}\left[4\mathcal{M}_{2}\left(\frac{1}{a}\right) - \frac{4}{a}\mathcal{M}_{3}\left(\frac{1}{a}\right) + \frac{1}{a^{2}}\mathcal{M}_{4}\left(\frac{1}{a}\right)\right], 
\label{eq:shift-2S} 
\end{equation} 
and 
\begin{equation} 
\delta E_{2P} = \frac{1}{24a^{5}}\mathcal{M}_{4}\left(\frac{1}{a}\right).
\label{eq:shift-2P} 
\end{equation} 
These expressions retain the complete dependence on $\omega a$, without a short or long range approximation. Their accuracy is nevertheless that of first order perturbation theory with nonrelativistic point Coulomb wave functions; corrections to the new interaction matrix elements must be included whenever the required precision resolves relativistic, radiative, or nuclear size effects.

For the Lamb-shift convention $\Delta E_{\mathrm{L}} = E_{2P_{1/2}} - E_{2S_{1/2}}$, the leading spin-independent correction becomes 
\begin{equation} 
\delta\Delta E_{\mathrm{L}} = -\frac{1}{2a^{3}}\mathcal{M}_{2}\left(\frac{1}{a}\right)+\frac{1}{2a^{4}}\mathcal{M}_{3}\left(\frac{1}{a}\right)-\frac{1}{12a^{5}}\mathcal{M}_{4}\left(\frac{1}{a}\right). 
\label{eq:atomic-Lamb-moments} 
\end{equation}
Introducing $u = 1 + (1-i)\omega a$, this result reduces to 
\begin{equation} 
\delta\Delta E_{\mathrm{L}} = \frac{\mathcal{C}a}{2}\left\{\operatorname{Re}(u^{-4})-\eta\varrho^{2}\operatorname{Im}\left(u^{-4}+2u^{-5}\right)\right\}+\mathcal{O}(\eta^{2}). 
\label{eq:atomic-Lamb-closed} 
\end{equation} 
The leading Coulomb like contribution $2\mathcal{C}\eta/r$ cancels between these levels because $\langle2S|r^{-1}|2S\rangle = \langle2P|r^{-1}|2P\rangle = 1/(4a)$. In this manner, the Lamb shift therefore probes the finite range dependence of the deformation instead of its leading $1/r$ term.

\begin{figure}[!htbp]
\centering
\makebox[\columnwidth][l]{
   \hspace*{-1.1cm}
   \includegraphics[width=1.1\columnwidth]{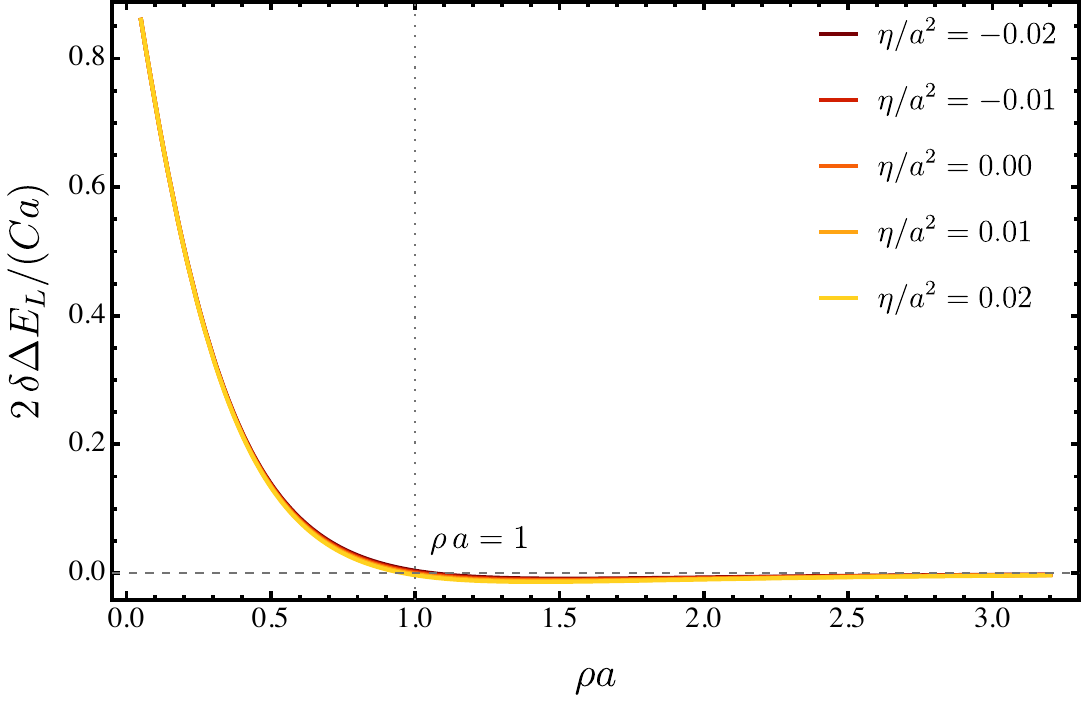}}
\caption{Dimensionless Lamb-shift response, $2\,\delta\Delta E_L/(Ca)$, as a function of the range parameter $\rho a$ for $\eta/a^2=\{-0.02,-0.01,0,0.01,0.02\}$.  The dashed horizontal line denotes zero response, while the dotted vertical line identifies the undeformed cancellation at $\rho a = 1$.}
\label{fig:hl-lamb-response}
\end{figure}

The range dependence of the Lamb shift response is displayed in Fig.~\ref{fig:hl-lamb-response}.  In the undeformed case the response vanishes at $\rho a = 1$.  A nonzero $\eta$ shifts the cancellation and slightly separates the curves on either side of it.  In this way, limits obtained near a zero of the response must retain the signed transition interval and account for higher order and matching uncertainties before being assigned physical significance.

When the interaction resolves the nucleus, the point-source potential must be replaced by 
\begin{equation} 
V_{\mathrm{HL}}^{\mathrm{ext}}(\mathbf{r}) = \int\mathrm{d}^{3}R\,\rho_{N}(\mathbf{R})V_{\mathrm{HL}}\!\left(|\mathbf{r}-\mathbf{R}|\right), \,\, \int\mathrm{d}^{3}R\,\rho_{N}(\mathbf{R})=1. 
\label{eq:finite-size-convolution} 
\end{equation}
Here, $\rho_{N}$ describes the spatial distribution of the nuclear charge associated with the new interaction, which need not coincide with the electromagnetic charge density. Nuclear-size corrections can matter even when $a$ exceeds the nuclear radius, depending on the interaction range and the required precision. This convolution accounts for the extended source but does not determine ultraviolet matching coefficients or justify extrapolating the derivative expansion beyond its cutoff.

For a transition $i\rightarrow f$, the correction at the retained order can be organized as 
\begin{equation} 
\delta\Delta E_{if}\equiv\delta E_{f}-\delta E_{i} = \frac{\mathfrak{g}_{N\ell}\Lambda_{\mathrm{HL}}^{2}}{4\pi}\left[\mathcal{A}_{if}(\varrho,a) + \eta\mathcal{B}_{if}(\varrho,a)\right], 
\label{eq:AB-decomposition} 
\end{equation} 
where the response functions follow from the corresponding matrix elements. To compare this prediction with data, we define the residual $d_{if}=\Delta E_{if}^{\mathrm{exp}} - \Delta E_{if}^{\mathrm{SM}}$. For Gaussian uncertainties, the allowed interval satisfies 
\begin{equation} 
\left|d_{if} - \delta\Delta E_{if}\right|\leq k_{\mathrm{CL}}\sigma_{if}, 
\label{eq:atomic-residual-condition} 
\end{equation} 
where $k_{\mathrm{CL}}$ specifies the two-sided confidence level and $\sigma_{if}$ includes experimental, theoretical, and external-input uncertainties with their correlations. Addition in quadrature is appropriate only for independent contributions.

A conservative bound on the magnitude follows by setting $\Delta E_{if}^{\mathrm{max}} = |d_{if}| + k_{\mathrm{CL}}\sigma_{if}$, which gives 
\begin{equation} 
|\mathfrak{g}_{N\ell}|\leq\frac{4\pi\Delta E_{if}^{\mathrm{max}}}{\Lambda_{\mathrm{HL}}^{2}\left|\mathcal{A}_{if}+\eta\mathcal{B}_{if}\right|}. 
\label{eq:coupling-bound}
\end{equation}
This symmetric bound is weaker than the signed interval in Eq.~\eqref{eq:atomic-residual-condition} when $d_{if}\neq0$. Nuclear radii and fundamental constants used in the Standard-Model prediction must be determined independently of the tested transition or fitted jointly with the new interaction \cite{Delaunay2023}. In particular, a charge radius extracted from a muonic Lamb shift under the Standard-Model hypothesis cannot be reused as an independent input to constrain an additional contribution to that same splitting.

A transition loses sensitivity at the retained order when $\mathcal{A}_{if} + \eta\mathcal{B}_{if} = 0$. Such zeros need not arise from interference between the two orders: Eq.~\eqref{eq:atomic-Lamb-closed} already vanishes at $\varrho a=1$ when $\eta=0$. Near a zero, neglected higher-order terms and matching uncertainties must be assessed before assigning an exclusion limit. Combining transitions can remove a cancellation only when their responses are independent and their coupling parameters are related by the assumed interaction model.

As an application, consider the Lamb shift of muonic helium-4, whose measured value is 
\begin{equation} 
\Delta E_{\mathrm{L}}^{\mathrm{exp}} = 1378.521(48)\,\mathrm{meV}. 
\label{eq:muonic-helium-experimental} 
\end{equation} 
The theoretical contributions compiled in Ref.~\cite{Pachucki2024} give 
\begin{equation} 
\begin{split}
& \Delta E_{\mathrm{L}}^{\mathrm{SM}}(r_{\alpha}) \\
&= \left[1668.491(7)+9.276(433)-106.209\left(\frac{r_{\alpha}}{\mathrm{fm}}\right)^{2}\right]\mathrm{meV}, 
\label{eq:muonic-helium-theory} 
\end{split}
\end{equation} 
where $r_{\alpha}$ is the electromagnetic charge radius of the alpha particle. For an illustrative independent-radius comparison, we use the electron-scattering value $r_{\alpha}=1.681(4)\,\mathrm{fm}$ quoted in Ref.~\cite{Krauth2021}, assuming that the new interaction is negligible in the reference scattering data.

Treating these inputs as independent Gaussian quantities and propagating the radius uncertainty to first order gives $d_{\mathrm{L}}\simeq0.875\,\mathrm{meV}$ and 
\begin{equation} 
\sigma_{\mathrm{L}} = \left[\sigma_{\mathrm{exp}}^{2}+\sigma_{\mathrm{QED}}^{2}+\sigma_{\mathrm{NS}}^{2}+\left(2C_{r}r_{\alpha}\sigma_{r_{\alpha}}\right)^{2}\right]^{1/2}\simeq1.493\,\mathrm{meV}, 
\label{eq:muonic-helium-uncertainty} 
\end{equation} 
with $C_{r}=106.209\,\mathrm{meV}\,\mathrm{fm}^{-2}$. At $95\%$ confidence, with $k_{\mathrm{CL}}=1.96$, this comparison permits \begin{equation} 
-2.05\,\mathrm{meV}\lesssim\delta\Delta E_{\mathrm{L}}\lesssim3.80\,\mathrm{meV}.
\label{eq:muonic-helium-interval} 
\end{equation} 
The corresponding conservative symmetric tolerance is $\Delta E_{\mu^{4}\mathrm{He}}^{\mathrm{max}}\simeq3.80\,\mathrm{meV}$, dominated by the independent-radius uncertainty. For this system, $a_{\mu^{4}\mathrm{He}} = 1/(2\alpha_{\mathrm{em}}\mu_{\mu\alpha})$, and Eqs.~\eqref{eq:atomic-Lamb-closed} and \eqref{eq:atomic-residual-condition} constrain the signed product $\mathfrak{g}_{\alpha\mu}$ wherever the point-source approximation is adequate. Resolving the nuclear distribution requires the extended-source matrix elements instead.

The same construction applies to electronic and muonic hydrogenic systems, with their respective reduced masses, nuclear inputs, and coupling products. Muonium probes the purely leptonic interaction between $e^{-}$ and $\mu^{+}$ without a composite nuclear source \cite{MuMASS2022}. Molecular ions and antiprotonic helium require few body wave functions and pair dependent interactions \cite{Salumbides2014,Germann2021}; for a normalized state $|\Psi_i\rangle$, the first order shift is 
\begin{equation} 
\delta E_i = \sum_{A<B}\int\mathrm{d}\tau\,|\Psi_i|^{2}V_{\mathrm{HL}}^{AB}(r_{AB}), 
\label{eq:few-body-shift} 
\end{equation}
where each $V_{\mathrm{HL}}^{AB}$ carries the coupling product and source structure appropriate to that pair. Joint limits from different systems require a specified relation among these couplings and a treatment of the shared experimental and theoretical uncertainties.


\section{Conclusion}
\label{sec:conclusion}

We constructed an isotropic higher spatial derivative deformation of Ho\v{r}ava--Lifshitz fermions, restricted to $|\eta|p^2\ll1$ below the effective theory cutoff. The spatial operator changed the dispersion relation without introducing additional frequency poles, since the action remained first order in time.

The mean field calculation retained particle and antiparticle excitations and determined their contribution to the finite temperature gap equation. Near a Fermi surface, the deformation changed the Fermi velocity and density of states, which entered exponentially in the weak coupling pairing scale. The ratio $T_c/\Delta_0 = e^{\gamma_E}/\pi$ survived under the usual constant density of states and weak coupling assumptions, while the $z=2$ gap equation remained cutoff dependent. The electromagnetic construction included the current and two photon contact vertices together with the condensate response. It identified the conditions for homogeneous superconductivity---a thermodynamically stable paired solution and positive static transverse stiffness---without assuming that every nonzero gap satisfied them.

For the specified $z=2$ exchange ansatz, the massive potential was exponentially damped and oscillatory. Its first order $1/r$ contribution described a resolved intermediate distance behavior when the cutoff hierarchy permitted it; extrapolation to the origin required ultraviolet matching. The massless potential instead contained a linearly growing term and did not admit the usual free scattering asymptotics. With the massive kernel, the density coupled Born cross section contained three distinct effects: the deformation of the group velocity, the exchange denominator, and the external spinor overlap. At fixed nonzero incident momentum, the nonzero screening scale made the forward limit and angular integral finite. Above that scale and below the effective cutoff, the leading orbital factor decreased as $q^{-8}$ and was multiplied by the derived Mott type spin factor.

The hydrogenic matrix elements retained the full interaction range dependence within first order perturbation theory. They also revealed a limitation of Lamb shift sensitivity: the leading Coulomb--like deformation canceled between $2S$ and $2P$, while finite range contributions remained. Their use in atomic spectroscopy required ordinary infrared kinetic energy and controlled source size corrections. For the independent radius inputs and Gaussian uncertainty treatment adopted for muonic helium--4, the additional Lamb shift contribution was restricted to approximately $[-2.05,3.80]\,\mathrm{meV}$ at $95\%$ confidence, with the external charge radius uncertainty dominating the comparison. 


\section*{Acknowledgments}
\hspace{0.5cm} A. A. Araújo Filho is supported by Conselho Nacional de Desenvolvimento Cient\'{\i}fico e Tecnol\'{o}gico (CNPq) -- [150223/2025-0].  The authors would like to thank BNAG event and J. Elias and I. P. Lobo for the fruitfull discussions in the quantum gravity Journal club. This work by A. F. Santos is partially supported by National Council for Scientific and Technological Development - CNPq project No. 312406/2023-1. K.E.L.F. thanks the CNPq for the financial support, Para\'iba State Research Support Foundation (FAPESQ) for the exchange program Para\'iba sem Fronteiras and the ENSEMBLE3 project, which is carried out within the 2.1 International Research Agendas programme of the Foundation for Polish Science co-financed by the European Union under the European Funds for Smart Economy 2021-2027 (FENG.02.01-IP.05-0044/24), project (MAB/2020/14), which is carried out within the International Research Agendas Programme (IRAP) of the Foundation for Polish Science co-financed by the European Union under the European Regional Development Fund and the Teaming Horizon 2020 programme (GA. No. 857543) of the European Commission and the project of the Minister of Science and Higher Education "Support for the activities of Centers of Excellence established in Poland under the Horizon 2020 program" under contract No. MEiN/2023/DIR/3797. M. Paganelly is supported by Conselho Nacional de Desenvolvimento Científico e Tecnológico (CNPq)
-- 152508/2024-4.

\section*{Data Availability Statement}

Data associated with this study consist of the analytical expressions and numerical figures presented in the manuscript. No additional dataset is required to reproduce the analytical results. The code used to generate some results and figures is available from the corresponding author upon reasonable request.

\bibliographystyle{apsrev4-2}
\bibliography{main}

\begin{thebibliography}{39}%
\makeatletter
\providecommand \@ifxundefined [1]{%
 \@ifx{#1\undefined}
}%
\providecommand \@ifnum [1]{%
 \ifnum #1\expandafter \@firstoftwo
 \else \expandafter \@secondoftwo
 \fi
}%
\providecommand \@ifx [1]{%
 \ifx #1\expandafter \@firstoftwo
 \else \expandafter \@secondoftwo
 \fi
}%
\providecommand \natexlab [1]{#1}%
\providecommand \enquote  [1]{``#1''}%
\providecommand \bibnamefont  [1]{#1}%
\providecommand \bibfnamefont [1]{#1}%
\providecommand \citenamefont [1]{#1}%
\providecommand \href@noop [0]{\@secondoftwo}%
\providecommand \href [0]{\begingroup \@sanitize@url \@href}%
\providecommand \@href[1]{\@@startlink{#1}\@@href}%
\providecommand \@@href[1]{\endgroup#1\@@endlink}%
\providecommand \@sanitize@url [0]{\catcode `\\12\catcode `\$12\catcode
  `\&12\catcode `\#12\catcode `\^12\catcode `\_12\catcode `\%12\relax}%
\providecommand \@@startlink[1]{}%
\providecommand \@@endlink[0]{}%
\providecommand \url  [0]{\begingroup\@sanitize@url \@url }%
\providecommand \@url [1]{\endgroup\@href {#1}{\urlprefix }}%
\providecommand \urlprefix  [0]{URL }%
\providecommand \Eprint [0]{\href }%
\providecommand \doibase [0]{https://doi.org/}%
\providecommand \selectlanguage [0]{\@gobble}%
\providecommand \bibinfo  [0]{\@secondoftwo}%
\providecommand \bibfield  [0]{\@secondoftwo}%
\providecommand \translation [1]{[#1]}%
\providecommand \BibitemOpen [0]{}%
\providecommand \bibitemStop [0]{}%
\providecommand \bibitemNoStop [0]{.\EOS\space}%
\providecommand \EOS [0]{\spacefactor3000\relax}%
\providecommand \BibitemShut  [1]{\csname bibitem#1\endcsname}%
\let\auto@bib@innerbib\@empty
\bibitem [{\citenamefont {Ho{\v{r}}ava}(2009)}]{Horava:2009uw}%
  \BibitemOpen
  \bibfield  {author} {\bibinfo {author} {\bibfnamefont {P.}~\bibnamefont
  {Ho{\v{r}}ava}},\ }\href {https://doi.org/10.1103/PhysRevD.79.084008}
  {\bibfield  {journal} {\bibinfo  {journal} {Physical Review D}\ }\textbf
  {\bibinfo {volume} {79}},\ \bibinfo {pages} {084008} (\bibinfo {year}
  {2009})},\ \Eprint {https://arxiv.org/abs/0901.3775} {arXiv:0901.3775
  [hep-th]} \BibitemShut {NoStop}%
\bibitem [{\citenamefont {Anselmi}(2009)}]{Anselmi:2008bq}%
  \BibitemOpen
  \bibfield  {author} {\bibinfo {author} {\bibfnamefont {D.}~\bibnamefont
  {Anselmi}},\ }\href {https://doi.org/10.1103/PhysRevD.79.025017} {\bibfield
  {journal} {\bibinfo  {journal} {Physical Review D}\ }\textbf {\bibinfo
  {volume} {79}},\ \bibinfo {pages} {025017} (\bibinfo {year} {2009})},\
  \Eprint {https://arxiv.org/abs/0808.3470} {arXiv:0808.3470 [hep-ph]}
  \BibitemShut {NoStop}%
\bibitem [{\citenamefont {Anselmi}\ and\ \citenamefont
  {Taiuti}(2010)}]{AnselmiTaiuti:2010}%
  \BibitemOpen
  \bibfield  {author} {\bibinfo {author} {\bibfnamefont {D.}~\bibnamefont
  {Anselmi}}\ and\ \bibinfo {author} {\bibfnamefont {M.}~\bibnamefont
  {Taiuti}},\ }\href {https://doi.org/10.1103/PhysRevD.81.085042} {\bibfield
  {journal} {\bibinfo  {journal} {Physical Review D}\ }\textbf {\bibinfo
  {volume} {81}},\ \bibinfo {pages} {085042} (\bibinfo {year} {2010})},\
  \Eprint {https://arxiv.org/abs/0912.0113} {arXiv:0912.0113 [hep-ph]}
  \BibitemShut {NoStop}%
\bibitem [{\citenamefont {Anselmi}(2008)}]{Anselmi:2008ry}%
  \BibitemOpen
  \bibfield  {author} {\bibinfo {author} {\bibfnamefont {D.}~\bibnamefont
  {Anselmi}},\ }\href {https://doi.org/10.1088/1126-6708/2008/02/051}
  {\bibfield  {journal} {\bibinfo  {journal} {JHEP}\ }\textbf {\bibinfo
  {volume} {02}},\ \bibinfo {pages} {051}},\ \Eprint
  {https://arxiv.org/abs/0801.1216} {arXiv:0801.1216 [hep-th]} \BibitemShut
  {NoStop}%
\bibitem [{\citenamefont {Barvinsky}\ \emph {et~al.}(2016)\citenamefont
  {Barvinsky}, \citenamefont {Blas}, \citenamefont {Herrero-Valea},
  \citenamefont {Sibiryakov},\ and\ \citenamefont
  {Steinwachs}}]{Barvinsky2016}%
  \BibitemOpen
  \bibfield  {author} {\bibinfo {author} {\bibfnamefont {A.~O.}\ \bibnamefont
  {Barvinsky}}, \bibinfo {author} {\bibfnamefont {D.}~\bibnamefont {Blas}},
  \bibinfo {author} {\bibfnamefont {M.}~\bibnamefont {Herrero-Valea}}, \bibinfo
  {author} {\bibfnamefont {S.~M.}\ \bibnamefont {Sibiryakov}},\ and\ \bibinfo
  {author} {\bibfnamefont {C.~F.}\ \bibnamefont {Steinwachs}},\ }\href
  {https://doi.org/10.1103/PhysRevD.93.064022} {\bibfield  {journal} {\bibinfo
  {journal} {Phys. Rev. D}\ }\textbf {\bibinfo {volume} {93}},\ \bibinfo
  {pages} {064022} (\bibinfo {year} {2016})},\ \Eprint
  {https://arxiv.org/abs/1512.02250} {arXiv:1512.02250 [hep-th]} \BibitemShut
  {NoStop}%
\bibitem [{\citenamefont {Barvinsky}\ \emph {et~al.}(2025)\citenamefont
  {Barvinsky}, \citenamefont {Kurov},\ and\ \citenamefont
  {Sibiryakov}}]{BarvinskyKurovSibiryakov2025}%
  \BibitemOpen
  \bibfield  {author} {\bibinfo {author} {\bibfnamefont {A.~O.}\ \bibnamefont
  {Barvinsky}}, \bibinfo {author} {\bibfnamefont {A.~V.}\ \bibnamefont
  {Kurov}},\ and\ \bibinfo {author} {\bibfnamefont {S.~M.}\ \bibnamefont
  {Sibiryakov}},\ }\href {https://doi.org/10.1103/PhysRevD.111.024030}
  {\bibfield  {journal} {\bibinfo  {journal} {Phys. Rev. D}\ }\textbf {\bibinfo
  {volume} {111}},\ \bibinfo {pages} {024030} (\bibinfo {year} {2025})},\
  \Eprint {https://arxiv.org/abs/2411.13574} {arXiv:2411.13574 [gr-qc]}
  \BibitemShut {NoStop}%
\bibitem [{\citenamefont {Montani}\ and\ \citenamefont
  {Schaposnik}(2012)}]{Montani:2012cu}%
  \BibitemOpen
  \bibfield  {author} {\bibinfo {author} {\bibfnamefont {H.}~\bibnamefont
  {Montani}}\ and\ \bibinfo {author} {\bibfnamefont {F.~A.}\ \bibnamefont
  {Schaposnik}},\ }\href {https://doi.org/10.1103/PhysRevD.86.065024}
  {\bibfield  {journal} {\bibinfo  {journal} {Physical Review D}\ }\textbf
  {\bibinfo {volume} {86}},\ \bibinfo {pages} {065024} (\bibinfo {year}
  {2012})},\ \Eprint {https://arxiv.org/abs/1206.1027} {arXiv:1206.1027
  [hep-th]} \BibitemShut {NoStop}%
\bibitem [{\citenamefont {Alexandre}\ and\ \citenamefont
  {Brister}(2013)}]{AlexandreBrister:2013}%
  \BibitemOpen
  \bibfield  {author} {\bibinfo {author} {\bibfnamefont {J.}~\bibnamefont
  {Alexandre}}\ and\ \bibinfo {author} {\bibfnamefont {J.}~\bibnamefont
  {Brister}},\ }\href {https://doi.org/10.1103/PhysRevD.88.065020} {\bibfield
  {journal} {\bibinfo  {journal} {Physical Review D}\ }\textbf {\bibinfo
  {volume} {88}},\ \bibinfo {pages} {065020} (\bibinfo {year} {2013})},\
  \Eprint {https://arxiv.org/abs/1307.7613} {arXiv:1307.7613 [hep-th]}
  \BibitemShut {NoStop}%
\bibitem [{\citenamefont {Alexandre}\ \emph {et~al.}(2012)\citenamefont
  {Alexandre}, \citenamefont {Brister},\ and\ \citenamefont
  {Houston}}]{AlexandreBristerHouston2012}%
  \BibitemOpen
  \bibfield  {author} {\bibinfo {author} {\bibfnamefont {J.}~\bibnamefont
  {Alexandre}}, \bibinfo {author} {\bibfnamefont {J.}~\bibnamefont {Brister}},\
  and\ \bibinfo {author} {\bibfnamefont {N.}~\bibnamefont {Houston}},\ }\href
  {https://doi.org/10.1103/PhysRevD.86.025030} {\bibfield  {journal} {\bibinfo
  {journal} {Phys. Rev. D}\ }\textbf {\bibinfo {volume} {86}},\ \bibinfo
  {pages} {025030} (\bibinfo {year} {2012})},\ \Eprint
  {https://arxiv.org/abs/1204.2246} {arXiv:1204.2246 [hep-ph]} \BibitemShut
  {NoStop}%
\bibitem [{\citenamefont {Iengo}\ \emph {et~al.}(2009)\citenamefont {Iengo},
  \citenamefont {Russo},\ and\ \citenamefont {Serone}}]{Iengo:2009ix}%
  \BibitemOpen
  \bibfield  {author} {\bibinfo {author} {\bibfnamefont {R.}~\bibnamefont
  {Iengo}}, \bibinfo {author} {\bibfnamefont {J.~G.}\ \bibnamefont {Russo}},\
  and\ \bibinfo {author} {\bibfnamefont {M.}~\bibnamefont {Serone}},\ }\href
  {https://doi.org/10.1088/1126-6708/2009/11/020} {\bibfield  {journal}
  {\bibinfo  {journal} {JHEP}\ }\textbf {\bibinfo {volume} {11}},\ \bibinfo
  {pages} {020}},\ \Eprint {https://arxiv.org/abs/0906.3477} {arXiv:0906.3477
  [hep-th]} \BibitemShut {NoStop}%
\bibitem [{\citenamefont {Kosteleck{\'y}}\ and\ \citenamefont
  {Russell}(2011)}]{KosteleckyRussell2011}%
  \BibitemOpen
  \bibfield  {author} {\bibinfo {author} {\bibfnamefont {V.~A.}\ \bibnamefont
  {Kosteleck{\'y}}}\ and\ \bibinfo {author} {\bibfnamefont {N.}~\bibnamefont
  {Russell}},\ }\href {https://doi.org/10.1103/RevModPhys.83.11} {\bibfield
  {journal} {\bibinfo  {journal} {Rev. Mod. Phys.}\ }\textbf {\bibinfo {volume}
  {83}},\ \bibinfo {pages} {11} (\bibinfo {year} {2011})},\ \bibinfo {note}
  {2026 edition: arXiv:0801.0287v19},\ \Eprint
  {https://arxiv.org/abs/0801.0287} {arXiv:0801.0287 [hep-ph]} \BibitemShut
  {NoStop}%
\bibitem [{\citenamefont {Myers}\ and\ \citenamefont
  {Pospelov}(2003)}]{MyersPospelov:2003}%
  \BibitemOpen
  \bibfield  {author} {\bibinfo {author} {\bibfnamefont {R.~C.}\ \bibnamefont
  {Myers}}\ and\ \bibinfo {author} {\bibfnamefont {M.}~\bibnamefont
  {Pospelov}},\ }\href {https://doi.org/10.1103/PhysRevLett.90.211601}
  {\bibfield  {journal} {\bibinfo  {journal} {Physical Review Letters}\
  }\textbf {\bibinfo {volume} {90}},\ \bibinfo {pages} {211601} (\bibinfo
  {year} {2003})},\ \Eprint {https://arxiv.org/abs/hep-ph/0301124}
  {arXiv:hep-ph/0301124} \BibitemShut {NoStop}%
\bibitem [{\citenamefont {Kosteleck{\'y}}\ and\ \citenamefont
  {Mewes}(2013)}]{KosteleckyMewes:2013}%
  \BibitemOpen
  \bibfield  {author} {\bibinfo {author} {\bibfnamefont {V.~A.}\ \bibnamefont
  {Kosteleck{\'y}}}\ and\ \bibinfo {author} {\bibfnamefont {M.}~\bibnamefont
  {Mewes}},\ }\href {https://doi.org/10.1103/PhysRevD.88.096006} {\bibfield
  {journal} {\bibinfo  {journal} {Physical Review D}\ }\textbf {\bibinfo
  {volume} {88}},\ \bibinfo {pages} {096006} (\bibinfo {year} {2013})},\
  \Eprint {https://arxiv.org/abs/1308.4973} {arXiv:1308.4973 [hep-ph]}
  \BibitemShut {NoStop}%
\bibitem [{\citenamefont {Maccione}\ \emph {et~al.}(2009)\citenamefont
  {Maccione}, \citenamefont {Taylor}, \citenamefont {Mattingly},\ and\
  \citenamefont {Liberati}}]{Maccione:2009ju}%
  \BibitemOpen
  \bibfield  {author} {\bibinfo {author} {\bibfnamefont {L.}~\bibnamefont
  {Maccione}}, \bibinfo {author} {\bibfnamefont {A.~M.}\ \bibnamefont
  {Taylor}}, \bibinfo {author} {\bibfnamefont {D.~M.}\ \bibnamefont
  {Mattingly}},\ and\ \bibinfo {author} {\bibfnamefont {S.}~\bibnamefont
  {Liberati}},\ }\href {https://doi.org/10.1088/1475-7516/2009/04/022}
  {\bibfield  {journal} {\bibinfo  {journal} {JCAP}\ }\textbf {\bibinfo
  {volume} {04}},\ \bibinfo {pages} {022}},\ \Eprint
  {https://arxiv.org/abs/0902.1756} {arXiv:0902.1756 [astro-ph.HE]}
  \BibitemShut {NoStop}%
\bibitem [{\citenamefont {Kosteleck{\'y}}\ and\ \citenamefont
  {Li}(2019)}]{KosteleckyLi:2019}%
  \BibitemOpen
  \bibfield  {author} {\bibinfo {author} {\bibfnamefont {V.~A.}\ \bibnamefont
  {Kosteleck{\'y}}}\ and\ \bibinfo {author} {\bibfnamefont {Z.}~\bibnamefont
  {Li}},\ }\href {https://doi.org/10.1103/PhysRevD.99.056016} {\bibfield
  {journal} {\bibinfo  {journal} {Physical Review D}\ }\textbf {\bibinfo
  {volume} {99}},\ \bibinfo {pages} {056016} (\bibinfo {year} {2019})},\
  \Eprint {https://arxiv.org/abs/1812.11672} {arXiv:1812.11672 [hep-ph]}
  \BibitemShut {NoStop}%
\bibitem [{\citenamefont {Bardeen}\ \emph {et~al.}(1957)\citenamefont
  {Bardeen}, \citenamefont {Cooper},\ and\ \citenamefont
  {Schrieffer}}]{Bardeen:1957mv}%
  \BibitemOpen
  \bibfield  {author} {\bibinfo {author} {\bibfnamefont {J.}~\bibnamefont
  {Bardeen}}, \bibinfo {author} {\bibfnamefont {L.~N.}\ \bibnamefont
  {Cooper}},\ and\ \bibinfo {author} {\bibfnamefont {J.~R.}\ \bibnamefont
  {Schrieffer}},\ }\href {https://doi.org/10.1103/PhysRev.108.1175} {\bibfield
  {journal} {\bibinfo  {journal} {Physical Review}\ }\textbf {\bibinfo {volume}
  {108}},\ \bibinfo {pages} {1175} (\bibinfo {year} {1957})}\BibitemShut
  {NoStop}%
\bibitem [{\citenamefont {Nambu}\ and\ \citenamefont
  {Jona-Lasinio}(1961)}]{NambuJonaLasinio1961}%
  \BibitemOpen
  \bibfield  {author} {\bibinfo {author} {\bibfnamefont {Y.}~\bibnamefont
  {Nambu}}\ and\ \bibinfo {author} {\bibfnamefont {G.}~\bibnamefont
  {Jona-Lasinio}},\ }\href {https://doi.org/10.1103/PhysRev.122.345} {\bibfield
   {journal} {\bibinfo  {journal} {Phys. Rev.}\ }\textbf {\bibinfo {volume}
  {122}},\ \bibinfo {pages} {345} (\bibinfo {year} {1961})}\BibitemShut
  {NoStop}%
\bibitem [{\citenamefont {Tong}\ \emph {et~al.}(2024)\citenamefont {Tong},
  \citenamefont {Wang}, \citenamefont {Zhang},\ and\ \citenamefont
  {Zhu}}]{Tong:2023krn}%
  \BibitemOpen
  \bibfield  {author} {\bibinfo {author} {\bibfnamefont {X.}~\bibnamefont
  {Tong}}, \bibinfo {author} {\bibfnamefont {Y.}~\bibnamefont {Wang}}, \bibinfo
  {author} {\bibfnamefont {C.}~\bibnamefont {Zhang}},\ and\ \bibinfo {author}
  {\bibfnamefont {Y.}~\bibnamefont {Zhu}},\ }\href
  {https://doi.org/10.1088/1475-7516/2024/04/022} {\bibfield  {journal}
  {\bibinfo  {journal} {JCAP}\ }\textbf {\bibinfo {volume} {04}},\ \bibinfo
  {pages} {022}},\ \Eprint {https://arxiv.org/abs/2304.09428} {arXiv:2304.09428
  [hep-th]} \BibitemShut {NoStop}%
\bibitem [{\citenamefont {Dhar}\ \emph {et~al.}(2009)\citenamefont {Dhar},
  \citenamefont {Mandal},\ and\ \citenamefont {Wadia}}]{DharMandalWadia2009}%
  \BibitemOpen
  \bibfield  {author} {\bibinfo {author} {\bibfnamefont {A.}~\bibnamefont
  {Dhar}}, \bibinfo {author} {\bibfnamefont {G.}~\bibnamefont {Mandal}},\ and\
  \bibinfo {author} {\bibfnamefont {S.~R.}\ \bibnamefont {Wadia}},\ }\href
  {https://doi.org/10.1103/PhysRevD.80.105018} {\bibfield  {journal} {\bibinfo
  {journal} {Phys. Rev. D}\ }\textbf {\bibinfo {volume} {80}},\ \bibinfo
  {pages} {105018} (\bibinfo {year} {2009})},\ \Eprint
  {https://arxiv.org/abs/0905.2928} {arXiv:0905.2928 [hep-th]} \BibitemShut
  {NoStop}%
\bibitem [{\citenamefont {Nambu}(1960)}]{Nambu:1960tm}%
  \BibitemOpen
  \bibfield  {author} {\bibinfo {author} {\bibfnamefont {Y.}~\bibnamefont
  {Nambu}},\ }\href {https://doi.org/10.1103/PhysRev.117.648} {\bibfield
  {journal} {\bibinfo  {journal} {Physical Review}\ }\textbf {\bibinfo {volume}
  {117}},\ \bibinfo {pages} {648} (\bibinfo {year} {1960})}\BibitemShut
  {NoStop}%
\bibitem [{\citenamefont {Guo}\ \emph {et~al.}(2012)\citenamefont {Guo},
  \citenamefont {Chien},\ and\ \citenamefont {He}}]{GuoChienHe:2012}%
  \BibitemOpen
  \bibfield  {author} {\bibinfo {author} {\bibfnamefont {H.}~\bibnamefont
  {Guo}}, \bibinfo {author} {\bibfnamefont {C.-C.}\ \bibnamefont {Chien}},\
  and\ \bibinfo {author} {\bibfnamefont {Y.}~\bibnamefont {He}},\ }\href
  {https://doi.org/10.1103/PhysRevD.85.074025} {\bibfield  {journal} {\bibinfo
  {journal} {Phys. Rev. D}\ }\textbf {\bibinfo {volume} {85}},\ \bibinfo
  {pages} {074025} (\bibinfo {year} {2012})},\ \Eprint
  {https://arxiv.org/abs/1202.5234} {arXiv:1202.5234 [math-ph]} \BibitemShut
  {NoStop}%
\bibitem [{\citenamefont {Touati}(2025)}]{Touati:2025}%
  \BibitemOpen
  \bibfield  {author} {\bibinfo {author} {\bibfnamefont {A.}~\bibnamefont
  {Touati}},\ }\href {https://doi.org/10.1016/j.physletb.2025.139598}
  {\bibfield  {journal} {\bibinfo  {journal} {Physics Letters B}\ }\textbf
  {\bibinfo {volume} {867}},\ \bibinfo {pages} {139598} (\bibinfo {year}
  {2025})},\ \Eprint {https://arxiv.org/abs/2410.15220} {arXiv:2410.15220
  [hep-th]} \BibitemShut {NoStop}%
\bibitem [{\citenamefont {Ara{\'u}jo~Filho}(2026)}]{AraujoFilho:2026yaj}%
  \BibitemOpen
  \bibfield  {author} {\bibinfo {author} {\bibfnamefont {A.~A.}\ \bibnamefont
  {Ara{\'u}jo~Filho}},\ }\href
  {https://doi.org/10.1140/epjc/s10052-026-16038-8} {\bibfield  {journal}
  {\bibinfo  {journal} {Eur. Phys. J. C}\ }\textbf {\bibinfo {volume} {86}},\
  \bibinfo {pages} {767} (\bibinfo {year} {2026})},\ \Eprint
  {https://arxiv.org/abs/2603.15959} {arXiv:2603.15959 [gr-qc]} \BibitemShut
  {NoStop}%
\bibitem [{\citenamefont {Ara{\'u}jo~Filho}\ \emph {et~al.}(2026)\citenamefont
  {Ara{\'u}jo~Filho}, \citenamefont {de~Farias}, \citenamefont {Passos},
  \citenamefont {Brito}, \citenamefont {{\"O}vg{\"u}n}, \citenamefont
  {Hassanabadi}, \citenamefont {Bezerra},\ and\ \citenamefont
  {Queiroz}}]{AraujoFilho:2026oqc}%
  \BibitemOpen
  \bibfield  {author} {\bibinfo {author} {\bibfnamefont {A.~A.}\ \bibnamefont
  {Ara{\'u}jo~Filho}}, \bibinfo {author} {\bibfnamefont {K.~E.~L.}\
  \bibnamefont {de~Farias}}, \bibinfo {author} {\bibfnamefont {E.}~\bibnamefont
  {Passos}}, \bibinfo {author} {\bibfnamefont {F.~A.}\ \bibnamefont {Brito}},
  \bibinfo {author} {\bibfnamefont {A.}~\bibnamefont {{\"O}vg{\"u}n}}, \bibinfo
  {author} {\bibfnamefont {H.}~\bibnamefont {Hassanabadi}}, \bibinfo {author}
  {\bibfnamefont {V.~B.}\ \bibnamefont {Bezerra}},\ and\ \bibinfo {author}
  {\bibfnamefont {A.~R.}\ \bibnamefont {Queiroz}},\ }\href
  {https://doi.org/10.1140/epjp/s13360-026-08048-y} {\bibfield  {journal}
  {\bibinfo  {journal} {Eur. Phys. J. Plus}\ }\textbf {\bibinfo {volume}
  {141}},\ \bibinfo {pages} {843} (\bibinfo {year} {2026})},\ \Eprint
  {https://arxiv.org/abs/2601.07102} {arXiv:2601.07102 [gr-qc]} \BibitemShut
  {NoStop}%
\bibitem [{\citenamefont {Born}(1926)}]{Born:1926}%
  \BibitemOpen
  \bibfield  {author} {\bibinfo {author} {\bibfnamefont {M.}~\bibnamefont
  {Born}},\ }\href {https://doi.org/10.1007/BF01397477} {\bibfield  {journal}
  {\bibinfo  {journal} {Zeitschrift f{\"u}r Physik}\ }\textbf {\bibinfo
  {volume} {37}},\ \bibinfo {pages} {863} (\bibinfo {year} {1926})}\BibitemShut
  {NoStop}%
\bibitem [{\citenamefont {Lippmann}\ and\ \citenamefont
  {Schwinger}(1950)}]{LippmannSchwinger:1950}%
  \BibitemOpen
  \bibfield  {author} {\bibinfo {author} {\bibfnamefont {B.~A.}\ \bibnamefont
  {Lippmann}}\ and\ \bibinfo {author} {\bibfnamefont {J.}~\bibnamefont
  {Schwinger}},\ }\href {https://doi.org/10.1103/PhysRev.79.469} {\bibfield
  {journal} {\bibinfo  {journal} {Physical Review}\ }\textbf {\bibinfo {volume}
  {79}},\ \bibinfo {pages} {469} (\bibinfo {year} {1950})}\BibitemShut
  {NoStop}%
\bibitem [{\citenamefont {Mott}(1929)}]{Mott:1929}%
  \BibitemOpen
  \bibfield  {author} {\bibinfo {author} {\bibfnamefont {N.~F.}\ \bibnamefont
  {Mott}},\ }\href {https://doi.org/10.1098/rspa.1929.0127} {\bibfield
  {journal} {\bibinfo  {journal} {Proceedings of the Royal Society of London.
  Series A}\ }\textbf {\bibinfo {volume} {124}},\ \bibinfo {pages} {425}
  (\bibinfo {year} {1929})}\BibitemShut {NoStop}%
\bibitem [{\citenamefont {Pohl}\ \emph {et~al.}(2010)\citenamefont {Pohl} \emph
  {et~al.}}]{Pohl2010}%
  \BibitemOpen
  \bibfield  {author} {\bibinfo {author} {\bibfnamefont {R.}~\bibnamefont
  {Pohl}} \emph {et~al.},\ }\href {https://doi.org/10.1038/nature09250}
  {\bibfield  {journal} {\bibinfo  {journal} {Nature}\ }\textbf {\bibinfo
  {volume} {466}},\ \bibinfo {pages} {213} (\bibinfo {year}
  {2010})}\BibitemShut {NoStop}%
\bibitem [{\citenamefont {Antognini}\ \emph {et~al.}(2013)\citenamefont
  {Antognini} \emph {et~al.}}]{Antognini2013}%
  \BibitemOpen
  \bibfield  {author} {\bibinfo {author} {\bibfnamefont {A.}~\bibnamefont
  {Antognini}} \emph {et~al.},\ }\href
  {https://doi.org/10.1126/science.1230016} {\bibfield  {journal} {\bibinfo
  {journal} {Science}\ }\textbf {\bibinfo {volume} {339}},\ \bibinfo {pages}
  {417} (\bibinfo {year} {2013})}\BibitemShut {NoStop}%
\bibitem [{\citenamefont {Pohl}\ \emph {et~al.}(2016)\citenamefont {Pohl} \emph
  {et~al.}}]{Pohl2016}%
  \BibitemOpen
  \bibfield  {author} {\bibinfo {author} {\bibfnamefont {R.}~\bibnamefont
  {Pohl}} \emph {et~al.},\ }\href {https://doi.org/10.1126/science.aaf2468}
  {\bibfield  {journal} {\bibinfo  {journal} {Science}\ }\textbf {\bibinfo
  {volume} {353}},\ \bibinfo {pages} {669} (\bibinfo {year}
  {2016})}\BibitemShut {NoStop}%
\bibitem [{\citenamefont {Krauth}\ \emph {et~al.}(2021)\citenamefont {Krauth},
  \citenamefont {Schuhmann}, \citenamefont {Ahmed} \emph
  {et~al.}}]{Krauth2021}%
  \BibitemOpen
  \bibfield  {author} {\bibinfo {author} {\bibfnamefont {J.~J.}\ \bibnamefont
  {Krauth}}, \bibinfo {author} {\bibfnamefont {K.}~\bibnamefont {Schuhmann}},
  \bibinfo {author} {\bibfnamefont {M.~A.}\ \bibnamefont {Ahmed}}, \emph
  {et~al.},\ }\href {https://doi.org/10.1038/s41586-021-03183-1} {\bibfield
  {journal} {\bibinfo  {journal} {Nature}\ }\textbf {\bibinfo {volume} {589}},\
  \bibinfo {pages} {527} (\bibinfo {year} {2021})}\BibitemShut {NoStop}%
\bibitem [{\citenamefont {Pachucki}\ \emph {et~al.}(2024)\citenamefont
  {Pachucki}, \citenamefont {Lensky}, \citenamefont {Hagelstein}, \citenamefont
  {Li~Muli}, \citenamefont {Bacca},\ and\ \citenamefont {Pohl}}]{Pachucki2024}%
  \BibitemOpen
  \bibfield  {author} {\bibinfo {author} {\bibfnamefont {K.}~\bibnamefont
  {Pachucki}}, \bibinfo {author} {\bibfnamefont {V.}~\bibnamefont {Lensky}},
  \bibinfo {author} {\bibfnamefont {F.}~\bibnamefont {Hagelstein}}, \bibinfo
  {author} {\bibfnamefont {S.~S.}\ \bibnamefont {Li~Muli}}, \bibinfo {author}
  {\bibfnamefont {S.}~\bibnamefont {Bacca}},\ and\ \bibinfo {author}
  {\bibfnamefont {R.}~\bibnamefont {Pohl}},\ }\href
  {https://doi.org/10.1103/RevModPhys.96.015001} {\bibfield  {journal}
  {\bibinfo  {journal} {Reviews of Modern Physics}\ }\textbf {\bibinfo {volume}
  {96}},\ \bibinfo {pages} {015001} (\bibinfo {year} {2024})}\BibitemShut
  {NoStop}%
\bibitem [{\citenamefont {Salumbides}\ \emph {et~al.}(2014)\citenamefont
  {Salumbides}, \citenamefont {Ubachs},\ and\ \citenamefont
  {Korobov}}]{Salumbides2014}%
  \BibitemOpen
  \bibfield  {author} {\bibinfo {author} {\bibfnamefont {E.~J.}\ \bibnamefont
  {Salumbides}}, \bibinfo {author} {\bibfnamefont {W.}~\bibnamefont {Ubachs}},\
  and\ \bibinfo {author} {\bibfnamefont {V.~I.}\ \bibnamefont {Korobov}},\
  }\href {https://doi.org/10.1016/j.jms.2014.04.003} {\bibfield  {journal}
  {\bibinfo  {journal} {Journal of Molecular Spectroscopy}\ }\textbf {\bibinfo
  {volume} {300}},\ \bibinfo {pages} {65} (\bibinfo {year} {2014})}\BibitemShut
  {NoStop}%
\bibitem [{\citenamefont {Germann}\ \emph {et~al.}(2021)\citenamefont
  {Germann}, \citenamefont {Patra}, \citenamefont {Karr}, \citenamefont
  {Hilico}, \citenamefont {Korobov}, \citenamefont {Salumbides}, \citenamefont
  {Eikema}, \citenamefont {Ubachs},\ and\ \citenamefont
  {Koelemeij}}]{Germann2021}%
  \BibitemOpen
  \bibfield  {author} {\bibinfo {author} {\bibfnamefont {M.}~\bibnamefont
  {Germann}}, \bibinfo {author} {\bibfnamefont {S.}~\bibnamefont {Patra}},
  \bibinfo {author} {\bibfnamefont {J.-P.}\ \bibnamefont {Karr}}, \bibinfo
  {author} {\bibfnamefont {L.}~\bibnamefont {Hilico}}, \bibinfo {author}
  {\bibfnamefont {V.~I.}\ \bibnamefont {Korobov}}, \bibinfo {author}
  {\bibfnamefont {E.~J.}\ \bibnamefont {Salumbides}}, \bibinfo {author}
  {\bibfnamefont {K.~S.~E.}\ \bibnamefont {Eikema}}, \bibinfo {author}
  {\bibfnamefont {W.}~\bibnamefont {Ubachs}},\ and\ \bibinfo {author}
  {\bibfnamefont {J.~C.~J.}\ \bibnamefont {Koelemeij}},\ }\href
  {https://doi.org/10.1103/PhysRevResearch.3.L022028} {\bibfield  {journal}
  {\bibinfo  {journal} {Physical Review Research}\ }\textbf {\bibinfo {volume}
  {3}},\ \bibinfo {pages} {L022028} (\bibinfo {year} {2021})}\BibitemShut
  {NoStop}%
\bibitem [{\citenamefont {Ohayon}\ \emph {et~al.}(2022)\citenamefont {Ohayon}
  \emph {et~al.}}]{MuMASS2022}%
  \BibitemOpen
  \bibfield  {author} {\bibinfo {author} {\bibfnamefont {B.}~\bibnamefont
  {Ohayon}} \emph {et~al.},\ }\href
  {https://doi.org/10.1103/PhysRevLett.128.011802} {\bibfield  {journal}
  {\bibinfo  {journal} {Physical Review Letters}\ }\textbf {\bibinfo {volume}
  {128}},\ \bibinfo {pages} {011802} (\bibinfo {year} {2022})}\BibitemShut
  {NoStop}%
\bibitem [{\citenamefont {Door}\ \emph {et~al.}(2025)\citenamefont {Door},
  \citenamefont {Yeh}, \citenamefont {Heinz} \emph {et~al.}}]{Door2025}%
  \BibitemOpen
  \bibfield  {author} {\bibinfo {author} {\bibfnamefont {M.}~\bibnamefont
  {Door}}, \bibinfo {author} {\bibfnamefont {C.-H.}\ \bibnamefont {Yeh}},
  \bibinfo {author} {\bibfnamefont {M.}~\bibnamefont {Heinz}}, \emph {et~al.},\
  }\href {https://doi.org/10.1103/PhysRevLett.134.063002} {\bibfield  {journal}
  {\bibinfo  {journal} {Phys. Rev. Lett.}\ }\textbf {\bibinfo {volume} {134}},\
  \bibinfo {pages} {063002} (\bibinfo {year} {2025})},\ \Eprint
  {https://arxiv.org/abs/2403.07792} {arXiv:2403.07792 [physics.atom-ph]}
  \BibitemShut {NoStop}%
\bibitem [{\citenamefont {Wilzewski}\ \emph {et~al.}(2025)\citenamefont
  {Wilzewski}, \citenamefont {Huber}, \citenamefont {Door} \emph
  {et~al.}}]{Wilzewski2025}%
  \BibitemOpen
  \bibfield  {author} {\bibinfo {author} {\bibfnamefont {A.}~\bibnamefont
  {Wilzewski}}, \bibinfo {author} {\bibfnamefont {L.~I.}\ \bibnamefont
  {Huber}}, \bibinfo {author} {\bibfnamefont {M.}~\bibnamefont {Door}}, \emph
  {et~al.},\ }\href {https://doi.org/10.1103/PhysRevLett.134.233002} {\bibfield
   {journal} {\bibinfo  {journal} {Phys. Rev. Lett.}\ }\textbf {\bibinfo
  {volume} {134}},\ \bibinfo {pages} {233002} (\bibinfo {year} {2025})},\
  \Eprint {https://arxiv.org/abs/2412.10277} {arXiv:2412.10277
  [physics.atom-ph]} \BibitemShut {NoStop}%
\bibitem [{\citenamefont {Delaunay}\ \emph {et~al.}(2023)\citenamefont
  {Delaunay}, \citenamefont {Karr}, \citenamefont {Kitahara}, \citenamefont
  {Koelemeij}, \citenamefont {Soreq},\ and\ \citenamefont
  {Zupan}}]{Delaunay2023}%
  \BibitemOpen
  \bibfield  {author} {\bibinfo {author} {\bibfnamefont {C.}~\bibnamefont
  {Delaunay}}, \bibinfo {author} {\bibfnamefont {J.-P.}\ \bibnamefont {Karr}},
  \bibinfo {author} {\bibfnamefont {T.}~\bibnamefont {Kitahara}}, \bibinfo
  {author} {\bibfnamefont {J.~C.~J.}\ \bibnamefont {Koelemeij}}, \bibinfo
  {author} {\bibfnamefont {Y.}~\bibnamefont {Soreq}},\ and\ \bibinfo {author}
  {\bibfnamefont {J.}~\bibnamefont {Zupan}},\ }\href
  {https://doi.org/10.1103/PhysRevLett.130.121801} {\bibfield  {journal}
  {\bibinfo  {journal} {Phys. Rev. Lett.}\ }\textbf {\bibinfo {volume} {130}},\
  \bibinfo {pages} {121801} (\bibinfo {year} {2023})},\ \bibinfo {note}
  {erratum: Phys. Rev. Lett. 134, 119901 (2025)},\ \Eprint
  {https://arxiv.org/abs/2210.10056} {arXiv:2210.10056 [hep-ph]} \BibitemShut
  {NoStop}%
\bibitem [{\citenamefont {Ara{\'u}jo~Filho}(2025)}]{AraujoFilho:2025fwd}%
  \BibitemOpen
  \bibfield  {author} {\bibinfo {author} {\bibfnamefont {A.~A.}\ \bibnamefont
  {Ara{\'u}jo~Filho}},\ }\href
  {https://doi.org/10.1140/epjc/s10052-025-14752-3} {\bibfield  {journal}
  {\bibinfo  {journal} {Eur. Phys. J. C}\ }\textbf {\bibinfo {volume} {85}},\
  \bibinfo {pages} {1002} (\bibinfo {year} {2025})},\ \Eprint
  {https://arxiv.org/abs/2504.19246} {arXiv:2504.19246 [gr-qc]} \BibitemShut
  {NoStop}%
\end{thebibliography}%

\end{document}